\documentclass[1p]{elsarticle}
\usepackage{amssymb,amsmath,epsfig} 
\usepackage{afterpage} 
\usepackage{array} 
\usepackage{amsfonts} 
\usepackage{hhline} 
\usepackage{subfigure}
 
\journal{Journal of Computational Physics} 
\begin{document} 
\begin{frontmatter} 
 
 
\title{Piecewise Parabolic Method on a Local Stencil for Magnetized Supersonic Turbulence Simulation} 
 
 
\author{Sergey~D.~Ustyugov, Mikhail~V.~Popov,}
\address{Keldysh Institute of Applied Mathematics, Miusskaya Sq.~4, 125047, Moscow, Russia} 
\author{Alexei~G.~Kritsuk$^*$, and Michael~L.~Norman}
\cortext[cor1]{Corresponding author}
\address{University of California, San Diego, 9500 Gilman Dr., La Jolla, CA 92093-0424, USA}
%\ead{ustyugs@keldysh.ru} 
 
\begin{abstract} 
Stable, accurate, divergence-free simulation of magnetized supersonic turbulence is a severe 
test of numerical MHD schemes and has been surprisingly difficult to achieve due to the range 
of flow conditions present. Here we present a new, higher order-accurate, low dissipation 
numerical method which requires no additional dissipation or local "fixes" for stable execution. 
We describe PPML, a local stencil variant of the popular PPM algorithm for solving the equations 
of compressible ideal magnetohydrodynamics. The principal difference between PPML and PPM is 
that cell interface states are evolved rather that reconstructed at every timestep, resulting 
in a compact stencil. Interface states are evolved using Riemann invariants containing all 
transverse derivative information. The conservation laws are updated in an unsplit fashion, 
making the scheme fully multidimensional. Divergence-free evolution of the magnetic field is 
maintained using the higher order-accurate constrained transport technique of Gardiner and 
Stone. The accuracy and stability of the scheme is documented against a bank of standard test 
problems drawn from the literature. The method is applied to numerical simulation of supersonic
MHD turbulence, which is important for many problems in astrophysics, including star formation 
in dark molecular clouds. PPML accurately reproduces in three-dimensions a transition to 
turbulence in highly compressible isothermal gas in a molecular cloud model. The low dissipation 
and wide spectral bandwidth of this method make it an ideal candidate for direct turbulence 
simulations.
\end{abstract} 
 
\begin{keyword} 
gas dynamics\sep magnetohydrodynamics\sep MHD turbulence\sep numerical methods\sep PPM\sep compact stencil 
 
\PACS 02.70.Bf\sep 47.11.Bc 
\end{keyword} 
\end{frontmatter} 
 
\section{Introduction} 
 
Piecewise Parabolic Method on a Local Stencil (PPML)~\cite{PPML1, PPML2} is a new numerical scheme 
developed for solving multidimensional compressible Euler (HD) and ideal magnetohydrodynamic (MHD) 
equations. The method is based on 
a piecewise parabolic approximation of variables inside individual grid cells. 
It is third order-accurate in space and it implies second order temporal accuracy. This method is 
an improvement over the popular PPM introduced by Colella and Woodward~\cite{Colella_84,Woodward_84} 
for nonmagnetized flows with strong shocks and extended by Dai and Woodward to compressible ideal 
magnetohydrodynamics~\cite{Dai_0}. PPM has been widely used in computational practice ever since, and
versions of the PPM gas dynamics scheme have been incorporated into a number of codes for astrophysical
applications~\cite{Woodward_07}.

The main new feature of PPML is the procedure for calculating interface values between adjacent cells. 
Instead of an interpolation procedure used in the original PPM formulation, which employs a four-point 
stencil, PPML relies on the information preserved from a previous time step. The required values 
are obtained via Riemann invariants transferred along the characteristic curves of the equations 
to cell boundaries using an approximate parabolic solution within a cell~\cite{Leer, Guinot, Sanders1, Sanders2, Sanders3}. To 
preserve the order of the scheme at local extrema, a monotonicity constraint is applied to these 
interface values~\cite{Suresh, Balsara, Rider, Sekora}. In a multidimensional case a monotonicity 
preserving method from~\cite{Barth} is additionally applied. The scheme is multidimensional as it 
keeps terms containing derivatives with respect to the tangential directions in the equations for 
wave amplitudes. This approach provides the left and right states for the Riemann problem based on 
multidimensional reconstruction. For the ideal MHD case in three dimensions, a zero-divergence 
constraint on the magnetic field is enforced by the use of the Stokes theorem (the so-called 
constrained transport approach~\cite{Evans}). An updated component of the electric field at a cell 
boundary is 
calculated by averaging the quantities obtained from given components of flux-vectors, taking into 
account a value of the electric field gradient and the information about the sign of the tangential 
velocity at that boundary~\cite{Gardiner}. 

PPML has been tested on a number HD and MHD problems that demonstrated the ability of the method 
to resolve discontinuous solutions without adding excessive dissipation. PPML provides a very 
accurate treatment of strong discontinuities, minimizes numerical dissipation of the kinetic and 
magnetic energy, and substantially improves the spectral bandwidth for compressible turbulence 
models compared to its predecessors~\cite{kritsuk09}.
 
In this paper we present a comprehensive description of the numerical method for 3D MHD simulations 
as well as results of numerical tests. 
The method has been substantially improved and remastered compared to its previous version 
presented in \cite{PPML1, PPML2}. The new PPML features in this paper include:
(i) an improved approach to maintaining zero divergence for the magnetic field following \cite{Gardiner}, 
see Section~5;
(ii) a monotonicity constraint proposed by Rider et al. \cite{Rider}, as described in Section 6;
(iii) an extended suite of test problems that includes a comparison with the FLASH3 MHD solver, see
Section~8; Finally, in Section~9 we illustrate the performance of PPML on a three-dimensional problem 
of highly compressible weakly magnetized forced turbulence with a sonic Mach number of 10. 
The problem of supersonic, super-Alfv\'enic turbulence with an isothermal equation of state 
proved to be a challenging regime for numerical modeling due to the presence of strong 
rarefactions and very high density contrasts in the flow. PPML scheme is perfectly stable
numerically on problems of this sort. We also briefly discuss the effects of the weak magnetic field 
on the spectral properties of MHD turbulence and obtain good correspondence between our numerical 
results and observed characteristics of supersonic turbulence in star forming molecular clouds.

\section{PPML description} 
 
Let us consider a homogeneous one-dimensional grid with the spacing $h$ and a function 
$q(x,t_0)\equiv q_0(x)$ defined on this grid at $t=t_0$. It is assumed that the function $q_0(x)$ 
can be approximated by a parabola inside every grid cell (Fig.~\ref{pic1}): 
\begin{figure}[!ht] 
\centering 
\includegraphics[width=85.329mm,height=25.875mm]{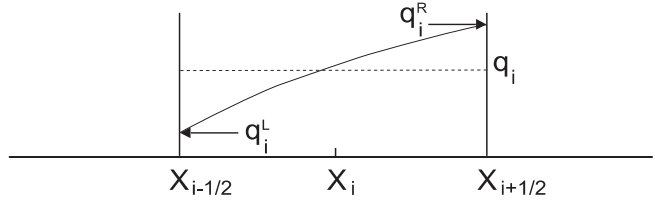} 
\caption{Approximation of $q(x)$ inside a cell.} 
\label{pic1} 
\end{figure} 
\begin{equation} 
\label{eq1} 
q_0(x)=q_i^L+\xi\left(\Delta q_i+q_i^{(6)}(1-\xi)\right), 
\end{equation}where 
$$ 
\xi=\left(x-x_{i-1/2}\right)h^{-1},\quad \Delta q_i=q_i^R-q_i^L, 
$$ 
$$ 
q_i^{(6)}=6\left(q_i-1/2\left(q_i^L+q_i^R\right)\right). 
$$ 
Function $q_i$ satisfies a condition 
$$ 
q_i=h^{-1}\int\limits_{x_{i-1/2}}^{x_{i+1/2}}q_0(x)\,dx. 
$$ 
 
\begin{figure}[t] 
\centering 
\includegraphics[width=58.795mm,height=40.19mm]{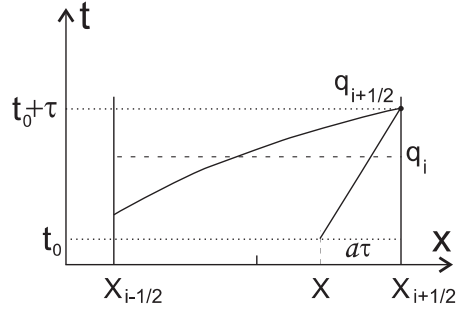} 
\caption{A characteristic inside a grid cell ($a>0$).} 
\label{pic2} 
\end{figure}

Let us consider a linear advection equation 
\begin{equation} 
\label{adv} 
\frac{\partial q(x,t)}{\partial t}+\frac{\partial F(x,t)}{\partial x}=0, 
\end{equation} 
where $F(x,t)=a\,q(x,t)$ and find its solution for the moment $t=t_0+\tau$. 
On a discrete grid we have a number of local Riemann problems which lead to 
some average states $q^*(x_{i+1/2},t)$ on every interface $x_{i+1/2}$ between 
the adjacent cells. The equation (\ref{adv}) has only one characteristic 
defined by $dx/dt=a$. To find a value $q_{i+1/2}$, for example, on the right 
boundary of a cell number $i$ at the time step $t=t_0+\tau$, we suggest to 
move along the characteristic line from the point $x_{i+1/2}$ back to the 
moment $t=t_0$ and use a value from some point on a previous parabola 
(Fig.~\ref{pic2}). 
 
Then for $a>0$ we have 
\begin{equation} 
\label{eq9} 
q_{i+1/2}(t_0+\tau)\equiv q_i^{\,R}(t_0+\tau)=q_i^{\,L}+\xi\left(\Delta q_i+q_i^{(6)}(1-\xi)\right), 
\end{equation} 
where 
\begin{equation} 
\label{xieq} 
\xi=\left(x-x_{i-1/2}\right)h^{-1}=\left(h-a\tau\right)h^{-1}=1-a\tau h^{-1}. 
\end{equation} 
All the values in the rhs of (\ref{eq9}) must be taken from a previous 
time step $t=t_0$. For $a<0$ a value $q_{i+1/2}$ is defined by a parabola 
in the cell number $i+1$: 
$$ 
q_{i+1/2}(t_0+\tau)\equiv q_{i+1}^{\,L}(t_0+\tau)=q_{i+1}^{\,L}+\xi\left(\Delta q_{i+1}+q_{i+1}^{(6)}(1-\xi)\right), 
$$ 
where $\xi=-a\tau h^{-1}$. 
 
In monotonic regions, where $q_{i+1/2}\in\left[q_i\ldots q_{i+1}\right]$, 
it is assumed that $q_i^{\,R}=q_{i+1}^{\,L}=q_{i+1/2}$ and $q_i^{\,L}=q_{i-1}^{\,R}=q_{i-1/2}$. 
In non-monotonic regions we must redefine $q_i^{\,L}$ and $q_i^{\,R}$. 
If $q_i$ is a local maximum or minimum, the interpolation function (\ref{eq1}) 
must be constant, i.e. $q_i^{\,L}=q_i^{\,R}=q_i$. 
If $q_i$ is too close to $q_i^{\,L}$ or $q_i^{\,R}$, the parabola (\ref{eq1}) 
can have an extremum inside the grid cell (it happens when $|\Delta q_i|<|q_i^{(6)}|$). 
In this case we must move this extremum to the boundary of the cell. These conditions 
can be written as  
\begin{equation} 
\label{eq4} 
q_i^{\,L}=q_i,\quad q_i^{\,R}=q_i,\quad\mbox{if}\;\left(q_i^{\,L}-q_i\right)\left(q_i-q_i^{\,R}\right)\le0  
\end{equation} 
and 
\begin{equation} 
\begin{array}{l} 
\label{eq5} 
q_i^{\,L}=3q_i-2q_i^{\,R},\quad\mbox{if}\;\;\Delta q_i\cdot q_i^{(6)}>\left(\Delta q_i\right)^2,\\ 
q_i^{\,R}=3q_i-2q_i^{\,L},\quad\mbox{if}\;\;\Delta q_i\cdot q_i^{(6)}<-\left(\Delta q_i\right)^2.\\ 
\end{array} 
\end{equation} 
If we know the function $q(x)$, we can compute its average for the interval 
\par\noindent $\left[x_{i+1/2}-y\ldots x_{i+1/2}\right]$ (for $y>0$): 
\begin{equation} 
\label{eq6} 
\overline{q}_{i+1/2}^{\,L}(y)=y^{-1}\int\limits_{x_{i+1/2}-y}^{x_{i+1/2}}q(x)\,dx 
=q_i^{\,R}-1/2\,y\,h^{-1}\left[\Delta q_i-\left(1-2/3\,y\,h^{-1}\right)q_i^{(6)}\right]. 
\end{equation} 
 
For $a>0$, the solution of (\ref{adv}) at time $t=t_0+\tau$ can be found by averaging 
over the interval 
$\left[x_{i+1/2}-a\tau\ldots x_{i+1/2}\right]$, i.e. $q^*(x_{i+1/2},
t_0+\tau)\equiv q_{i+1/2}^{\,L}=\overline{q}_{i+1/2}^{\,L}(a\tau)$. 
For $a<0$ the zone of influence is $\left[x_{i+1/2}\ldots x_{i+1/2}+a\tau\right]$. 
In this case $q^*(x_{i+1/2},t_0+\tau)\equiv q_{i+1/2}^{\,R}=\overline{q}_{i+1/2}^{\,R}(-a\tau)$, where 
\begin{multline} 
\label{eq7} 
\overline{q}_{i+1/2}^{\,R}(y)=y^{-1}\int\limits_{x_{i+1/2}}^{x_{i+1/2}+y}q(x)\,dx\\ 
=q_{i+1}^{\,L}+1/2\,y\,h^{-1}\left[\Delta q_{i+1}+\left(1-2/3\,y\,h^{-1}\right)q_{i+1}^{(6)}\right],\quad y>0. 
\end{multline} 
The flux on the interface can be computed as 
\begin{equation} 
\label{eq8} 
F_{i+1/2}=a^+q_{i+1/2}^{\,L}+a^-q_{i+1/2}^{\,R}, 
\end{equation} 
where $a^+=\max(a,0)=\left(a+|a|\right)/2$, $a^-=\min(a,0)=\left(a-|a|\right)/2$. 
We can use an arbitrary value for $q_{i+1/2}^{\,L}$ if $a<0$, and for $q_{i+1/2}^{\,R}$ if $a>0$. 
\vspace{12pt} 
 
\section{The governing equations} 
 
Let us consider the ideal MHD equations in 3D in the following form: 
\begin{equation} 
\label{eqxx1} 
\partial_t{\bf U}+\partial_x{\bf F}+\partial_y{\bf G}+\partial_z{\bf H}=0. 
\end{equation} 
Here ${\bf U}$ is a vector of eight conservative variables, ${\bf F}$, ${\bf G}$ and ${\bf H}$ are the fluxes:
\begin{equation} 
\label{es5} 
{\bf U}=\left(\rho,\rho\,u,\rho\,v,\rho\,w,B_x,B_y,B_z,E\right)^T, 
\end{equation} 
\begin{multline} 
\label{es6} 
{\bf F}=\left(\rho\,u,\rho\,u^2+\overline p-B_x^2,\rho uv-B_xB_y,\rho uw-B_xB_z,0,\right.\\ 
\bigl.uB_y-vB_x,uB_z-wB_x,u(E+\overline p)-B_x(uB_x+vB_y+wB_z)\bigr)^T, 
\end{multline} 
\begin{multline} 
\label{es6b} 
{\bf G}=\left(\rho\,v,\rho uv-B_xB_y,\rho\,v^2+\overline p-B_y^2,\rho vw-B_yB_z,vB_x-uB_y,\right.\\ 
\bigl.0,vB_z-wB_y,v(E+\overline p)-B_y(uB_x+vB_y+wB_z)\bigr)^T, 
\end{multline} 
\begin{multline} 
\label{es6bH} 
{\bf H}=\left(\rho\,w,\rho uw-B_xB_z,\rho vw-B_yB_z,\rho\,w^2+\overline p-B_z^2,wB_x-uB_z,\right.\\ 
\bigl.wB_y-vB_z,0,w(E+\overline p)-B_z(uB_x+vB_y+wB_z)\bigr)^T, 
\end{multline} 
where $\rho$ is the density; $u$, $v$ and $w$ are the velocity components; $B_x$, $B_y$ and $B_z$ are the magnetic field components; $E$ is the total energy and $\overline p$ is the total pressure: 
\begin{equation*} 
\label{es1a} 
\overline p=p+\frac{\bf BB}2. 
\end{equation*} 
We have included the factor $1/\sqrt{4\pi}$ in the definition of ${\bf B}$. 
An equation for the total energy and an ideal gas equation of state are 
\begin{equation*} 
\label{es2} 
E=\rho\,\varepsilon+\frac{\rho{\bf vv}}2+\frac{{\bf BB}}2, 
\end{equation*} 
\begin{equation*} 
\label{es2a} 
p=(\gamma-1)\,\rho\,\varepsilon, 
\end{equation*} 
where $\gamma$ - the adiabatic index, $\varepsilon$ - the specific internal energy. If we denote 
\begin{equation*} 
\label{es8} 
(b_x,b_y,b_z)=\frac1{\sqrt{\rho}}\left(B_x,B_y,B_z\right),\quad b^2=b_x^2+b_y^2+b_z^2, 
\end{equation*} 
we can write the sound velocity $c$, Alfv\'en velocity $c_a$, the fast and the slow magneto-acoustic velocities $c_{f,\,s}$ as 
\begin{equation*} 
\label{es9} 
c=\sqrt{\frac{\gamma p}{\rho}}, 
\end{equation*} 
\begin{equation*} 
\label{es10} 
c_a=|b_x|, 
\end{equation*} 
\begin{equation*} 
\label{es11} 
c_{f,\,s}=\left[\frac12\left(c^{\,2}+b^{\,2}\right)\pm\frac12\sqrt{\left(c^{\,2}+b^{\,2}\right)^{\,2}-4c^{\,2}b_x^{\,2}}\right]^{1/2}. 
\end{equation*} 
We will also deal with a non-conservative form of the MHD equations: 
\begin{equation} 
\label{es11a} 
\partial_t{\bf V}+A\,\partial_x{\bf V}+B\,\partial_y{\bf V}+C\,\partial_y{\bf V}=0, 
\end{equation} 
where 
\begin{equation} 
\label{es11b} 
{\bf V}=\left(\rho,u,v,w,B_x,B_y,B_z,p\,\right)^T. 
\end{equation} 
 
The matrices $A$, $B$ and $C$ can be computed using Jacobians of (\ref{eqxx1}) and a transition matrix   
\begin{equation*} 
\label{eqxx3a}  
M=\frac{\partial{\,\bf U}}{\partial{\,\bf V}}. 
\end{equation*} 
For example, the Jacoby matrix $A$ is 
\begin{equation} 
\label{es11f}  
A=M^{-1}\,\frac{\partial{\,\bf F}}{\partial{\,\bf U}}\,M=\left( 
\begin{array}{cccccccc} 
u&\rho&0&0&0&0&0&0\\ 
0&u&0&0&0&B_y/\rho&B_z/\rho&1/\rho\\ 
0&0&u&0&0&-B_x/\rho&0&0\\ 
0&0&0&u&0&0&-B_x/\rho&0\\ 
0&0&0&0&u&0&0&0\\ 
0&B_y&-B_x&0&0&u&0&0\\ 
0&B_z&0&-B_x&0&0&u&0\\ 
0&\gamma p&0&0&0&0&0&u\\ 
\end{array} 
\right). 
\end{equation} 
The corresponding eigenvalues are 
\begin{equation*} 
\label{es16} 
\lambda^{\,1,\,8}_{\,x}=u\pm c_f,\quad \lambda^{\,2,\,7}_{\,x}=u\pm c_a,\quad \lambda^{\,3,\,6}_{\,x}=u\pm c_s,\quad \lambda^{\,4,\,5}_{\,x}=u. 
\end{equation*} 
$\lambda^{\,1,\,8}_{\,x}$ represent a pair of fast magneto-acoustic waves, $\lambda^{\,2,\,7}_{\,x}$ -- 
a pair of Alfv\'en waves, $\lambda^{\,3,\,6}_{\,x}$ -- a pair of slow magneto-acoustic waves, 
$\lambda^{\,4}_{\,x}$ -- an entropy wave, $\lambda^{\,5}_{\,x}$ -- a magnetic-flux wave. 
The eigenvectors of the Jacobians could have singularities at the points of degeneracy of 
the eigenvalues since the MHD equations are nonstrictly hyperbolic. 
To avoid those, Brio and Wu~\cite{Brio_88} suggested a scaled version of the eigenvectors 
that comes from defining
\begin{equation*} 
\label{es13} 
(\beta_y,\beta_z)=\left\{ 
 \begin{aligned} 
  \frac{(B_y,B_z)}{\sqrt{B_y^{\,2}+B_z^{\,2}}}\quad\mbox{if}\;B_y^{\,2}+B_z^{\,2}\ne0,\\ 
  \left(\frac1{\sqrt{2}},\frac1{\sqrt{2}}\right)\;\mbox{otherwise},\\ 
 \end{aligned} 
\right. 
\end{equation*} 
\begin{equation*} 
\label{es14} 
(\alpha_f,\alpha_s)=\left\{ 
 \begin{aligned} 
  \frac{\left(\sqrt{\vphantom{c_f^2}c^{\,2}-c_s^{\,2}},\sqrt{c_f^{\,2}-c^{\,2}}\right)}{\sqrt{c_f^{\,2}-c_s^{\,2}}}\quad\mbox{if}\;B_y^{\,2}+B_z^{\,2}\ne0\;\mbox{or}\;\gamma p\ne B_x^{\,2},\\ 
  \left(\frac1{\sqrt{2}},\frac1{\sqrt{2}}\right)\;\mbox{otherwise.}\\ 
 \end{aligned} 
\right. 
\end{equation*} 
Thus the left and the right eigenvectors are 
\begin{multline*} 
%\label{es23} 
{\bf l}^{\,1,\,8}_{\,x}=\left(\vphantom{\frac12}0,\pm\frac{\alpha_f\,c_f}{2c^{\,2}},\mp\frac{\alpha_s}{2c^{\,2}}\,c_s\,\beta_y\,{\rm sgn}B_x,\mp\frac{\alpha_s}{2c^{\,2}}\,c_s\,\beta_z\,{\rm sgn}B_x,\right.\\ 
\left.0,\frac{\alpha_s}{2\sqrt{\rho}\,c}\,\beta_y,\frac{\alpha_s}{2\sqrt{\rho}\,c}\,\beta_z,\frac{\alpha_f}{2\rho\,c^{\,2}}\vphantom{\frac12}\right), 
\end{multline*} 
\begin{multline*} 
%\label{es24} 
{\bf r}^{\,1,\,8}_{\,x}=\left(\vphantom{A^{A^A}}\rho\,\alpha_f,\pm\alpha_f\,c_f,\mp\alpha_s\,c_s\,\beta_y\,{\rm sgn}B_x,\mp\alpha_s\,c_s\,\beta_z\,{\rm sgn}B_x,\right.\\ 
\left.0,\alpha_s\sqrt{\rho}\,c\,\beta_y,\alpha_s\sqrt{\rho}\,c\,\beta_z,\alpha_f\gamma\,p\vphantom{A^{A^A}}\right)^T, 
\end{multline*} 
\begin{equation*} 
%\label{es25} 
{\bf l}^{\,2,\,7}_{\,x}=\left(0,0,-\frac{\beta_z}{\sqrt{2}}\,{\rm sgn}B_x,\frac{\beta_y}{\sqrt{2}}\,{\rm sgn}B_x,0,\pm\frac{\beta_z}{\sqrt{2\rho}},\mp\frac{\beta_y}{\sqrt{2\rho}},0\right),
\end{equation*} 
\begin{equation*} 
%\label{es26} 
{\bf r}^{\,2,\,7}_{\,x}=\left(0,0,-\frac{\beta_z}{\sqrt{2}}\,{\rm sgn}B_x,\frac{\beta_y}{\sqrt{2}}\,{\rm sgn}B_x,0,\pm\sqrt{\frac{\rho}2}\,\beta_z,\mp\sqrt{\frac{\rho}2}\,\beta_y,0\right)^T, 
\end{equation*} 
\begin{multline*} 
%\label{es27} 
{\bf l}^{\,3,\,6}_{\,x}=\left(\vphantom{\frac12}0,\pm\frac{\alpha_s\,c_s}{2c^{\,2}},\pm\frac{\alpha_f}{2c^{\,2}}\,c_f\,\beta_y\,{\rm sgn}B_x,\pm\frac{\alpha_f}{2c^{\,2}}\,c_f\,\beta_z\,{\rm sgn}B_x,\right.\\ 
\left.0,-\frac{\alpha_f}{2\sqrt{\rho}\,c}\,\beta_y,-\frac{\alpha_f}{2\sqrt{\rho}\,c}\,\beta_z,\frac{\alpha_s}{2\rho\,c^{\,2}}\vphantom{\frac12}\right), 
\end{multline*} 
\begin{multline*} 
%\label{es28} 
{\bf r}^{\,3,\,6}_{\,x}=\left(\Bigr.\rho\,\alpha_s,\pm\alpha_s\,c_s,\pm\alpha_f\,c_f\,\beta_y\,{\rm sgn}B_x,\pm\alpha_f\,c_f\,\beta_z\,{\rm sgn}B_x,\right.\\ 
\left.0,-\alpha_f\sqrt{\rho}\,c\,\beta_y,-\alpha_f\sqrt{\rho}\,c\,\beta_z,\alpha_s\gamma\,p\vphantom{A^{A^A}}\right)^T, 
\end{multline*} 
\begin{equation*} 
%\label{es29} 
{\bf l}^{\,4}_{\,x}=\left(1,0,0,0,0,0,0,-\frac1{c^{,2}}\right), 
\end{equation*} 
\begin{equation*} 
%\label{es30} 
{\bf r}^{\,4}_{\,x}=\left(1,0,0,0,0,0,0,0\right)^T, 
\end{equation*} 
\begin{equation*} 
%\label{es30a} 
{\bf l}^{\,5}_{\,x}=\left(0,0,0,0,1,0,0,0\right), 
\end{equation*} 
\begin{equation*} 
%\label{es30b} 
{\bf r}^{\,5}_{\,x}=\left(0,0,0,0,1,0,0,0\right)^T. 
\end{equation*} 
 
\section{A numerical scheme} 
 
To solve (\ref{eqxx1}) we apply a conservative difference scheme: 
\begin{multline} 
\label{eqxx2} 
{\bf U}^{n+1}_{i,j,k}={\bf U}^n_{i,j,k}-\frac{\tau}{\Delta x}\left({\bf F}^{\,n+1/2}_{\,i+1/2,\,j,\,k}-{\bf F}^{\,n+1/2}_{\,i-1/2,\,j,\,k}\right)-\\ 
\frac{\tau}{\Delta y}\left({\bf G}^{\,n+1/2}_{\,i,\,j+1/2,\,k}-{\bf G}^{\,n+1/2}_{\,i,\,j-1/2,\,k}\right)- 
\frac{\tau}{\Delta z}\left({\bf H}^{\,n+1/2}_{\,i,\,j,\,k+1/2}-{\bf H}^{\,n+1/2}_{\,i,\,j,\,k-1/2}\right). 
\end{multline} 
Half-integer indices such as $i+1/2$ denote the boundaries of grid cells, 
half-integer time index $n+1/2$ means that we use the averaged values of 
the fluxes over a time step $\tau$ in order to get a second-order temporal 
accuracy. 

The solution inside every grid cell is approximated by a parabola along any 
Cartesian grid axis. The boundary values for each parabola are determined from 
a conservation property of Riemann invariants that remain constant along the 
characteristics of the initial linearized system of equations. Parabolae must 
be built using the primitive variables (\ref{es11b}), so we need to consider 
a non-conservative form of the MHD equations (\ref{es11a}). 
 
We can expand $A$, $B$ and $C$ in (\ref{es11a}) into their eigenvectors. 
For example, for the $x$-direction: 
\begin{equation} 
\label{eqxx4} 
A=R_{\,x}\,\Lambda_{\,x}\,L_{\,x}, 
\end{equation} 
where $R_{\,x}$ is a matrix with columns filled by the right eigenvectors 
${\bf r}_{\,x}^{\,p}$ ($p=1,\ldots,8$), $L_x=R_x^{-1}$ is an inverse matrix, 
with rows filled by the left eigenvectors ${\bf l}_{\,x}^{\,p}$. 
$\Lambda_{\,x}$ is a diagonal matrix of the eigenvalues: $(\Lambda_{\,x})_{ij}=0$ 
for $i\ne j$, $(\Lambda_{\,x})_{ij}=\lambda^{\,p}$ for $i=j=p\,$. 
 
To construct piecewise parabolae for every time step, one needs to define 
the states on the cell edges and the states at their centers. 
For simplicity, let us further consider a 1D case: 
\begin{equation} 
\label{eqxx7} 
\partial_t{\bf V}(x,t)+A\,\partial_x{\bf V}(x,t)=0. 
\end{equation} 
Inserting (\ref{eqxx4}) into (\ref{eqxx7}) and multiplying by the $L$ matrix from the left, 
we arrive at 
\begin{equation} 
\label{eqxx8} 
L\,\partial_{\,t}{\bf V}+\Lambda L\,\partial_{\,x}{\bf V}=0. 
\end{equation} 
Let us expand a vector ${\bf V}(x,t)$ into the local basis of the right eigenvectors 
${\bf r}^{\,p}$, which are fixed in every cell: 
\begin{equation} 
\label{eqxx9} 
{\bf V}(x,t)=\sum_p\alpha^{\,p}(x,t)\,{\bf r}^{\,p}. 
\end{equation} 
Inserting (\ref{eqxx9}) into (\ref{eqxx8}), we arrive at 
\begin{equation} 
\label{eqxx10} 
\partial_{\,t}\,\alpha^{\,p}+\lambda^{\,p}\,\partial_{\,x}\,\alpha^{\,p}=0,\quad p=1,\ldots,8. 
\end{equation} 
The equations (\ref{eqxx10}) mean that the coefficients $\alpha^{\,p}\,(x,t)$ in the expansion 
(\ref{eqxx9}) (the wave amplitudes) must be constant along the characteristics $x^{\,p}\,(t)$: 
$$ 
\frac{d\,x^{\,p}}{d\,t}=\lambda^{\,p}, 
$$  
i.e. $\alpha^{\,p}\,(x,t)$ are Riemann invariants. A value of Riemann invariant on 
the boundary of a cell ($x=x_{i+1/2}$) at the moment $t+\tau$ could be computed using 
its value at the moment $t$ as 
\begin{equation} 
\label{eqxx11} 
\alpha^{\,p}\,(x_{i+1/2},t+\tau)=\alpha^{\,p}\,(x_{i+1/2}-\lambda^{\,p}\,\tau,t). 
\end{equation} 
\begin{figure}[t] 
\centering 
\includegraphics[width=84.172mm,height=41.25mm]{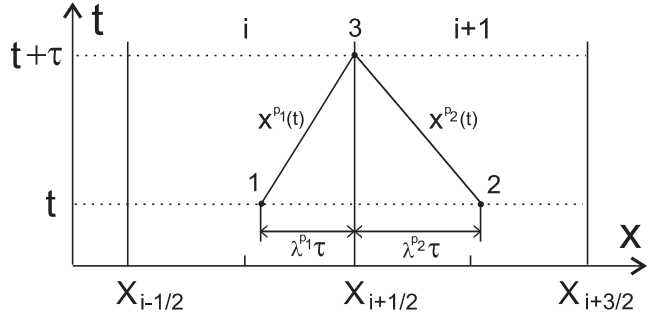} 
\caption{The characteristics in the adjacent cells for $\lambda^{p_{\,1}}>0$ and $\lambda^{p_{\,2}}<0$.} 
\label{pix1} 
\end{figure} 
 
Fig.~\ref{pix1} shows two adjacent cells $i$ and $i+1$. 
The characteristics in the cell $i$ have index $p_{\,1}$, in the cell $i+1$ -- 
index $p_{\,2}$. One of the characteristics $x^{\,p_{\,1}}\,(t)$ with the eigenvalue 
$\lambda^{p_{\,1}}>0$ is shown in the cell $i$, another one $x^{\,p_{\,2}}\,(t)$ with 
the eigenvalue $\lambda^{p_{\,2}}<0$ is shown in the cell $i+1$. According to (\ref{eqxx11}) 
the amplitude of a wave at point~3, which propagates inside the cell $i$ along the 
characteristic $x^{\,p_{\,1}}\,(t)$ with the eigenvalue $\lambda^{p_{\,1}}$, is equal 
to its value at point~1. In the same way the amplitude of a wave at point~3, which 
propagates inside the cell $i+1$ along the characteristic $x^{\,p_{\,2}}\,(t)$, is 
equal to its value at point~2.  
 
The state at point~3, which is computed according to (\ref{eqxx9}) by summation 
with respect to all the eigenvectors, fixed in cell $i$, with $\lambda^{p_{\,1}}>0$ 
will be on the left side of the interface. Let ${\bf V}^L$ denote this value. 
Similarly, let ${\bf V}^R$ denote the state at point~3 on the right side of the 
interface which is computed by summation with respect to all the eigenvectors fixed 
in cell $i+1$, with $\lambda^{p_{\,2}}<0$. 
 
The amplitudes of waves 
$\alpha^{\,p}\,(x^p,t)$ at the point $x^p=x_{i+1/2}-\lambda^{\,p}\,\tau$ 
in (\ref{eqxx11}), which influence the right boundary of cell $i$ ($\lambda^{\,p}>0$), 
can be computed by multiplying the expansion (\ref{eqxx9}) by the left eigenvectors, 
fixed in cell $i$: 
\begin{equation} 
\label{eqxx14a} 
\alpha^{\,p}\,(x^p,t)={\bf l}^{\,p}\:{\bf V}(x^p,t),\quad\lambda^{\,p}>0, 
\end{equation} 
where
\begin{equation} 
{\bf V}(x^p)\equiv\frac{1}{x_{i+1/2}-x^p}\int^{x_{i+1/2}}_{x^p}{\bf V}(x)dx. 
\end{equation} 

We can use arbitrary values for the wave amplitudes for $\lambda^{\,p}<0$ because these 
waves have no effect on the right boundary of the cell and will be omitted in 
the sum (\ref{eqxx9}). For convenience let them be 
\begin{equation} 
\label{eqxx14a2} 
\alpha^{\,p}\,(x^p,t)={\bf l}^{\,p}\:{\bf V}(x_i,t),\quad\lambda^{\,p}<0. 
\end{equation} 
With the help of $\Theta$-function we can rearrange (\ref{eqxx14a})-(\ref{eqxx14a2}) as 
\begin{equation} 
\label{eqxx20} 
\alpha^{\,p}\,(x^p,t)=\Theta(\lambda^{\,p})\Bigl({\bf  l}^{\,p}\,{\bf V}(x^p,t)\Bigr)+\left(1-\Theta(\lambda^{\,p})\right)\Bigl({\bf  l}^{\,p}\,{\bf V}(x_i,t)\Bigr). 
\end{equation} 
Multiplying (\ref{eqxx20}) by ${\bf r}^{\,p}$ and summing over all $p$ such as 
$\lambda^{\,p}>0$ according to (\ref{eqxx9}), after some simple manipulations 
we obtain the boundary value at time $t+\tau$: 
\begin{equation} 
\label{eqxx21n} 
{\bf V}^{L}(x_{i+1/2},t+\tau)={\bf V}(x_i,t)+\sum_{p\;(\lambda^{\,p}>0)}{\bf  r}^{\,p}\left[{\bf  l}^{\,p}\Bigl({\bf V}(x^p,t)-{\bf V}(x_i,t)\Bigr)\right]. 
\end{equation} 
 
If we consider cell $i+1$ and waves with $\lambda^{\,p}<0$ we will obtain a 
similar expression for the value ${\bf V}^R$ on the right side of the interface at 
time $t+\tau$: 
\begin{equation} 
\label{eqxx22n} 
{\bf V}^{R}(x_{i+1/2},t+\tau)={\bf V}(x_{i+1},t)+\sum_{p\;(\lambda^{\,p}<0)}
{\bf  r}^{\,p}\left[{\bf  l}^{\,p}\Bigl({\bf V}(x^p,t)-{\bf V}(x_{i+1},t)\Bigr)\right]. 
\end{equation} 
 
The left and the right eigenvectors in (\ref{eqxx21n})-(\ref{eqxx22n}) are 
fixed in every cell. To compute them we can use a state from any point $x$ 
inside the cell - it has been shown by numerical experiments that this choice 
has no influence on the solution. We suggest using the values of states in 
the centres of the cells, i.e. ${\bf l}^{\,p}={\bf l}^{\,p}({\bf V}(x_i,t))$. 
 
\begin{figure}[!ht] 
\centering 
\includegraphics[width=84.172mm,height=41.25mm]{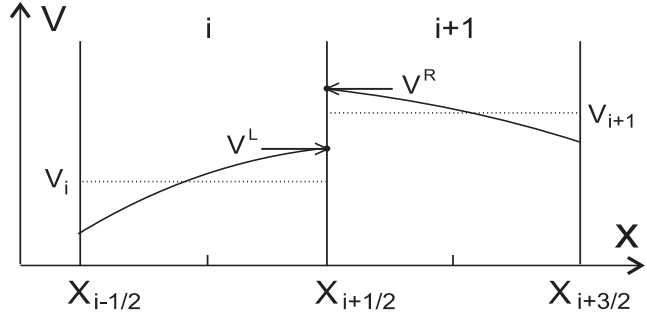} 
\par\caption{Approximation of ${\bf V}(x)$ in the adjacent cells.} 
\label{pix2} 
\end{figure} 
 
A possible approximation of the component $V(x)$ of the vector-function 
${\bf V}(x,t)$ inside grid cells $i$ and $i+1$ at some time step is shown 
in Fig.~\ref{pix2}. The arrows point to the values on the left and the 
right sides of the interface, note that $V^L\ne V^R$. 
The dotted lines are the average values of $V(x)$: 
$$ 
V_i=\frac1{\Delta x}\int\limits_{x_{i-1/2}}^{x_{i+1/2}}V(x)\,d\,x. 
$$ 
 
In order to solve the 3D problem we split the initial set (\ref{es11a}) 
by the space variables and solve the 1D equations separately for the 
$x$-, $y$- and $z$-directions. However, in this case we have an additional 
change in the quantities because of the fluxes in the orthogonal directions. 
\begin{figure}[!h] 
\centering 
\includegraphics[width=45mm,height=45mm]{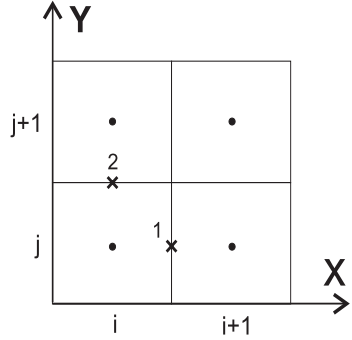} 
\par\caption{A 2D mesh.} 
\label{pix4} 
\end{figure} 
For example, a flux in the $y$-direction will affect the quantities at point~1 
between cells $(i,j)$ and $(i+1,j)$ (see Fig.~\ref{pix4}). To obtain the 
correct result, we can solve equation (\ref{es11a}) in the $x$-direction 
considering the terms $B\,\partial_y{\bf V}$ and $C\,\partial_z{\bf V}$ as 
sources. Then instead of (\ref{eqxx10}) we arrive at 
\begin{equation} 
\label{eqxx27b} 
\partial_{\,t}\,\alpha^{\,p}+\lambda^{\,p}\,\partial_{\,x}\,\alpha^{\,p}=-D^{\,p},\quad p=1,\ldots,8, 
\end{equation} 
where $D^{\,p}$ are the components of the vector
\begin{equation}
\label{eqxx27c} 
{\bf D}=L_x\left(B\:\partial_{\,y}{\bf V}\right)+L_x\left(C\:\partial_{\,z}{\bf V}\right). 
\end{equation} 
The components of the partial derivatives in (\ref{eqxx27c}) can be calculated as  
\begin{equation} 
\label{Deq} 
\partial_{\,y}V^p=\frac{V^p_{i,\,j+1/2,\,k}-V^p_{i,\,j-1/2,\,k}}{\Delta y}. 
\end{equation} 
The solution of (\ref{eqxx27b}) can be obtained from a Taylor series expansion of 
$\alpha^{\,p}\,(x,t)$ near the boundary $x=x_{i+1/2}$. It 
is similar to (\ref{eqxx11}) but has the additional term 
\begin{equation} 
\label{eqxx27e} 
\alpha^{\,p}\,(x_{i+1/2},t+\tau)=\alpha^{\,p}\,(x_{i+1/2}-\lambda^{\,p}\,\tau,t)-D^{\,p}(t)\,\frac{\tau}2. 
\end{equation} 
Then instead of (\ref{eqxx21n})-(\ref{eqxx22n}) we arrive at 
\begin{equation} 
\label{eqxx21b} 
{\bf V}^{L}(x_{i+1/2},t+\tau)={\bf V}(x_i,t) 
+\sum_{p\;(\lambda^{\,p}>0)}{\bf  r}^{\,p}\left[{\bf  l}^{\,p}\Bigl({\bf V}(x^p,t)-{\bf V}(x_i,t)-D^{\,p}(t)\,\frac{\tau}2\Bigr)\right], 
\end{equation} 
\begin{equation} 
\label{eqxx22b} 
{\bf V}^{R}(x_{i+1/2},t+\tau)={\bf V}(x_{i+1},t) 
+\sum_{p\;(\lambda^{\,p}<0)}{\bf  r}^{\,p}\left[{\bf  l}^{\,p}\Bigl({\bf V}(x^p,t)-{\bf V}(x_{i+1},t)-D^{\,p}(t)\,\frac{\tau}2\Bigr)\right]. 
\end{equation} 
In (\ref{eqxx27e})-(\ref{eqxx22b}) we have omitted indices $j$ and $k$. 
The state at point~2 on Fig.~\ref{pix4} is computed in a similar way 
considering the $x$-derivative as a source. 

To obtain states ${\bf V}_{i+1/2}$ in PPML, we solve the Riemann problem for the ${\bf V}^L$ 
and ${\bf V}^R$ states on every interface using, e.g., the Roe solver~\cite{Roe} or the HLLD 
solver~\cite{HLLD}:
\begin{equation}
\label{Riem}
{\bf V}_{i+1/2}=R\left({\bf V}^L,{\bf V}^R\right),
\end{equation}
where $R$ is the Riemann solver. 
Note, that in the original PPM, the states ${\bf V}_{i+1/2}$ are obtained through monotonic
interpolation \cite{Leer}. 
We then apply a monotonicity- and extrema-preserving procedure proposed by Rider et al. \cite{Rider} 
to the values of ${\bf V}_{i+1/2}$, as described in Section~\ref{mono}. 
Finally, we modify the resulting interface states with the standard PPM 
monotonicity procedure (\ref{eq4})-(\ref{eq5}). 
 
So far we have obtained the boundary values of piecewise parabolae at time 
$t+\tau$ in each grid cell. Now we need to define the fluxes to compute the 
new central states. Again we can use the Roe solver~\cite{Roe} or the HLLD 
solver~\cite{HLLD} with ${\bf V}^{L}$ and ${\bf V}^{R}$ from 
(\ref{eqxx21b})-(\ref{eqxx22b}) but in this case the numerical scheme will 
have the first order of temporal accuracy. To design a second order scheme, 
we must average the amplitudes $\alpha^{\,p}\,(x,t)$ over the zones of 
influence. 
 
\begin{figure}[!ht] 
\centering 
\includegraphics[width=56.062mm,height=41.25mm]{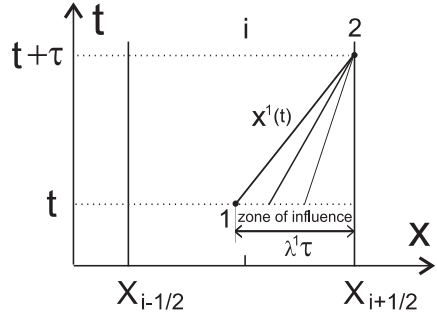} 
\par\caption{Characteristics inside a grid cell.} 
\label{pix3} 
\end{figure} 
 
Fig.~\ref{pix3} shows a set of characteristics corresponding to waves with 
$\lambda^{\,p}>0$. The characteristic $x^{\,1}\,(t)$ has the maximum eigenvalue 
$\lambda^{\,1}$. Point~1 is the point of intersection between this 
characteristic and the piecewise parabola at time $t$. Obviously, only the 
zone between the interface $x=x_{i+1/2}$ and point~1 affects the left boundary 
state at point~2. 
 
If we consider a wave which propagates along the characteristic with 
$\lambda^{\,p}>0$ inside cell $i$, its averaged amplitude on the interface 
$x=x_{i+1/2}$ at time $t+\tau$ can be calculated as 
\begin{equation} 
\label{eqxx14} 
\overline{\mathstrut\alpha}^{\,p}_{i+1/2}=\frac1{\lambda^{\,p}\,\tau}\int\limits_{x_{i+1/2}-\lambda^{\,p}\,\tau}^{x_{i+1/2}}\alpha^{\,p}\,(x)\,d\,x,\quad\lambda^{\,p}>0. 
\end{equation} 
Multiplying the expansion (\ref{eqxx9}) by the left eigenvectors, fixed in cell $i$, 
will yield 
\begin{equation} 
\label{eqxx14aa} 
\alpha^{\,p}\,(x)={\bf l}^{\,p}\:{\bf V}(x),\quad\lambda^{\,p}>0. 
\end{equation} 
Inserting (\ref{eqxx14aa}) into (\ref{eqxx14}) and removing the factor ${\bf l}^{\,p}$ 
from the integral, we arrive at 
\begin{equation*} 
\label{eqxx18} 
\overline{\mathstrut\alpha}^{\,p}_{i+1/2}={\bf l}^{\,p}\overline{\vphantom{V^L}\,{\bf V}}^{\;L,\,p}_{i+1/2}\,, 
\end{equation*} 
\begin{equation} 
\label{eqxx18a} 
\overline{\vphantom{V^L}\,{\bf V}}^{\;L,\,p}_{i+1/2}=\frac1{\lambda^{\,p}\,\tau}\int\limits_{x_{i+1/2}-\lambda^{\,p}\,\tau}^{x_{i+1/2}}{\bf V}(x)\,d\,x,\quad\lambda^{\,p}>0. 
\end{equation} 
After that we can arrive at 
\begin{equation} 
\label{eqxx21z} 
\overline{\vphantom{V^L}\,{\bf V}}^{\;L}=\overline{\vphantom{V^L}{\bf V}}^{\;L,\,1}_{i+1/2}+\sum_{p\;(\lambda^{\,p}>0)}{\bf  r}^{\,p}\left[{\bf  l}^{\,p}\Bigl(\overline{\vphantom{V^L}{\bf V}}^{\;L,\,p}_{i+1/2}-\overline{\vphantom{V^L}{\bf V}}^{\;L,\,1}_{i+1/2}-D^{\,p}\,\frac{\tau}2\Bigr)\right] 
\end{equation} 
which is similar to (\ref{eqxx21b}). Here $\overline{\vphantom{V^L}{\bf V}}^{\;L,\,1}_{i+1/2}$ 
is the averaged by formula (\ref{eqxx18a}) solution ${\bf V}(x)$ at time $t$ over the zone 
of influence of the wave in cell $i$, corresponding to the maximum eigenvalue ($\lambda^{\,1}>0$).  
 
For cell $i+1$ and the negative eigenvalues we arrive at 
\begin{equation} 
\label{eqxx22z} 
\overline{\vphantom{V^R}\,{\bf V}}^{\,R}=\overline{\vphantom{V^L}{\bf V}}^{\;R,\,1}_{i+1/2}+\sum_{p\;(\lambda^{\,p}<0)}{\bf  r}^{\,p}\left[{\bf  l}^{\,p}\Bigl(\overline{\vphantom{V^L}{\bf V}}^{\;R,\,p}_{i+1/2}-\overline{\vphantom{V^L}{\bf V}}^{\;R,\,1}_{i+1/2}-D^{\,p}\,\frac{\tau}2\Bigr)\right], 
\end{equation} 
\begin{equation} 
\label{eqxx18aneg} 
\overline{\vphantom{V^R}\,{\bf V}}^{\;R,\,p}_{i+1/2}=\frac1{|\lambda^{\,p}|\,\tau}\int\limits_{x_{i+1/2}}^{x_{i+1/2}-\lambda^{\,p}\,\tau}{\bf V}(x)\,d\,x,\quad\lambda^{\,p}<0. 
\end{equation} 
To get the second order of accuracy for time we must compute fluxes 
on every interface solving the Riemann problem with 
$\overline{\vphantom{V^L}\,{\bf V}}^{\;L}$ and $\overline{\vphantom{V^R}\,{\bf V}}^{\,R}$ 
from (\ref{eqxx21z})-(\ref{eqxx22z}). Note that the values of 
components of the integrals (\ref{eqxx18a}) and (\ref{eqxx18aneg}) 
can be computed as (\ref{eq6}) and (\ref{eq7}), respectively. 
 
\begin{figure}[t] 
  \centering 
  \includegraphics[width=60mm,height=60mm]{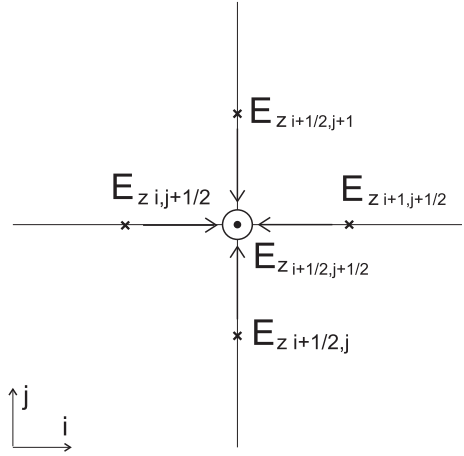} 
  \caption{\small Computation of the component $E_z$ at a node.} 
  \label{node} 
\end{figure} 

The time step $\tau$ in (\ref{eqxx2}) is obtained from the Courant condition
\begin{equation} 
\tau=\sigma\min_{i,j,k}\left\{\frac{\Delta x}{\left|u_{i,j,k}\right|+c_{f\;i,j,k}^{\,x}},\frac{\Delta y}{\left|v_{i,j,k}\right|+c_{f\;i,j,k}^{\,y}},\frac{\Delta z}{\left|w_{i,j,k}\right|+c_{f\;i,j,k}^{\,z}}\right\},
\end{equation} 
where $\sigma$ is the Courant number and $c_f^x$, $c_f^y$, and $c_f^z$ are the 
fast magneto-acoustic velocities along the coordinate directions.

\section{Zero divergence constraint for the magnetic field} 
\label{field} 
 
Our numerical method must provide numerical solutions that satisfy a condition 
\begin{equation} 
\label{es3} 
{\rm div}\,{\bf B}=0. 
\end{equation} 
There are several approaches to this problem in the 
literature~\cite{Gardiner,Toth,Powell,Brackbill,Evans,Balsara_S}. 
In our numerical scheme we used an unsplit Godunov method for ideal MHD with 
a constrained transport developed in~\cite{Gardiner} that is based 
on the Stokes theorem 
\begin{equation} 
\label{es43} 
\frac{\partial{\bf\,B}}{\partial\,t}=-\nabla\times{\bf E}. 
\end{equation} 
Calculated from (\ref{es43}) the magnetic field obviously satisfies condition 
(\ref{es3}). To apply the Stokes theorem in 3D case, we define the magnetic field
components on the cell faces and the electric field components on the
cell edges. The algorithm exploits the fact that some components of the fluxes 
${\bf F}$, ${\bf G}$ and ${\bf H}$ are actually the components of the electric field.  
 
For simplicity let us consider a 2D case. Thus the sixth component of the 
flux ${\bf F}$ is the $z$-component of the electric field reversed in sign, 
which correspond to the left and right boundaries of a cell. The fifth 
component of the flux ${\bf G}$ is the $z$-component of electric field, 
which correspond to the top and bottom boundaries of a cell. Computing Taylor 
series for these components near the nodes of the mesh and averaging them, we 
obtain the components of the electric field at the nodes (Fig.~\ref{node}). 
 
\begin{multline*} 
%\label{es45} 
E_{z\;i+1/2,j+1/2}=\frac14\left(E^{left}_{z\;i+1/2,j+1/2}+E^{right}_{z\;i+1/2,j+1/2}\right.\\
\left.+E^{top}_{z\;i+1/2,j+1/2}+E^{bottom}_{z\;i+1/2,j+1/2}\right), 
\end{multline*} 
where for example 
\begin{equation*} 
\label{es45b} 
E^{left}_{z\;i+1/2,j+1/2}=E_{z\;i,j+1/2}+\left.\frac{\partial E_z}{\partial x}\right|_{i,j+1/2}\frac{\Delta x}2, 
\end{equation*} 
\begin{equation*} 
\label{es45c} 
\left.\frac{\partial E_z}{\partial x}\right|_{i,j+1/2}=\left\{ 
\begin{aligned} 
\left(\frac{\partial E_z}{\partial x}\right)_{i,j},\quad v_{i,j+1/2}>0,\\ 
\left(\frac{\partial E_z}{\partial x}\right)_{i,j+1},\quad v_{i,j+1/2}<0,\\ 
\frac12\left[\left(\frac{\partial E_z}{\partial x}\right)_{i,j}+\left(\frac{\partial E_z}{\partial x}\right)_{i,j+1}\right],\quad v_{i,j+1/2}=0.\\ 
\end{aligned} 
\right. 
\end{equation*} 
These values of $E_{z\;i+1/2,j+1/2}$ are used in a discrete version of (\ref{es43}): 
\begin{equation*} 
\label{es46} 
B_{x\;i+1/2,j}^{n+1}=B_{x\;i+1/2,j}^n-\frac{\tau}{\Delta y}\left(\vphantom{A^{A^A}}E_{z\;i+1/2,j+1/2}-E_{z\;i+1/2,j-1/2}\right), 
\end{equation*} 
\begin{equation*} 
\label{es47} 
B_{y\;i,j+1/2}^{n+1}=B_{y\;i,j+1/2}^n+\frac{\tau}{\Delta x}\left(\vphantom{A^{A^A}}E_{z\;i+1/2,j+1/2}-E_{z\;i-1/2,j+1/2}\right). 
\end{equation*} 
 
Updated components of magnetic field at the center of $(i,j)$-cell are computed by averaging: 
\begin{equation*} 
\label{es48} 
B_{x\;i,j}^{n+1}=\frac12\left(\vphantom{A^{A^A}}B_{x\;i-1/2,j}^{n+1}+B_{x\;i+1/2,j}^{n+1}\right), 
\end{equation*} 
\begin{equation*} 
\label{es49} 
B_{y\;i,j}^{n+1}=\frac12\left(\vphantom{A^{A^A}}B_{y\;i,j-1/2}^{n+1}+B_{y\;i,j+1/2}^{n+1}\right). 
\end{equation*} 
 
The magnetic field computed this way will automatically satisfy (\ref{es3}). To make 
sure that this is indeed the case, ${\rm div}{\bf B}$ should be approximated by the following 
expression 
\begin{multline*} 
%\label{es50} 
\left({\rm div}\,{\bf B}\right)_{i+1/2,j+1/2}=\frac1{2\,\Delta x}\left(\vphantom{A^{A^A}}B_{x\;i+1,j}+B_{x\;i+1,j+1}-B_{x\;i,j}-B_{x\;i,j+1}\right)\\ 
+\frac1{2\,\Delta y}\left(\vphantom{A^{A^A}}B_{y\;i,j+1}+B_{y\;i+1,j+1}-B_{y\;i,j}-B_{y\;i+1,j}\right). 
\end{multline*} 
 
\section{Monotonicity constraints} 
\label{mono} 
 
A standard PPM monotonicity preserving procedure (\ref{eq4})-(\ref{eq5}) is insufficient for
the ideal MHD case. We need to use additional procedures to suppress spurious oscillations. 
 
{\bf Procedure~1.} To keep a solution monotonic without reducing the order of the scheme 
and to preserve all the local extrema in MHD simulations we can follow a number of approaches 
described in the literature~\cite{Suresh,Balsara,Rider,Sekora}. In our method we rely on a 
Piecewise Parabolic Accurate Monotonicity- and Extrema-Preserving procedure described 
in~\cite{Rider}. 

We compute the wave amplitudes $\alpha_i^p$ for the central states ${\bf V}_i$ 
and $\alpha_{i\pm1/2}^p$ for the interface states ${\bf V}_{i\pm1/2}$.
We then calculate new values for wave amplitudes of the interface states:
$$
\alpha_{i\pm1/2}^{p,\,*}={\rm median}\left(\alpha_i^p,\alpha_{i\pm1/2}^p,\alpha_{i\pm1}^p\right)
$$
and
$$
\alpha_{i\pm1/2}^{p,\,**}={\rm median}\left(\alpha_i^p,\alpha_{i\pm1/2}^{p,\,*},3\alpha_i^p-2\alpha_{i\mp1/2}^{p,\,*}\right),
$$
where the ${\rm median}$ function is defined in a usual way
$$
{\rm median}\left(a,b,c\right)=a+{\rm minmod}\left(b-a,c-a\right)
$$
through the ${\rm minmod}$ function
$$
{\rm minmod}\left(a,b\right)=\frac12\left({\rm sign}(a)+{\rm sign}(b)\right)\min\left(|a|,|b|\right).
$$
If $\alpha_{i\pm1/2}^{p,\,**}=\alpha_{i\pm1/2}^p$ for all $p$, 
the procedure is completed. 
Otherwise, we compute a set of fifth-order WENO interface values $\alpha_{i\pm1/2,*}^p$, 
using Algorithm 2.2.4 from~\cite{Rider}, and check for a local extremum. 
If $\alpha_{i\pm1/2}^{p,\,**}=\alpha_i^p$ for all $p$, we apply Algorithm 2.1.2, 4(b), 
otherwise the region is monotonic but too steep to be approximated with the values
obtained from the Riemann solver (\ref{Riem}) and we apply Algorithm 2.1.2, 4(c).
 
{\bf Procedure~2.} 
To keep a solution monotonic in a multidimensional case, we employ a method proposed
in~\cite{Barth}.

Let us consider a 2D case for simplicity. A parabola 
that approximates a solution along the $x$-axis for every component $V(x)$ of a state ${\bf V}(x,t)$ 
at some point in time can be defined as 
\begin{equation*} 
\label{es54} 
V(x)=V_{\,i,j}+\phi(V)\left[s_{\,i,j}\,(x-x_i)+\frac{\sigma_{\,i,j}}2\left((x-x_i)^2-\frac{\Delta x^2}{12}\right)\right], 
\end{equation*} 
where 
\begin{equation*} 
\label{es55} 
s_{\,i,j}=\frac{V_{i+1/2,j}-V_{i-1/2,j}}{\Delta x},\quad \sigma_{\,i,j}=6\,\frac{V_{i+1/2,j}-2V_{i,j}+V_{i-1/2,j}}{\Delta x^2}. 
\end{equation*} 
As a limiting function $\phi(V)$ we can use that described in~\cite{Barth}: 
\begin{multline} 
\label{es56} 
\phi(V)=\min\left(1,\frac{|V_{i,j}-\max(V_{l,m})|}{|V_{i,j}-\max(V_{i-1/2,j},V_{i+1/2,j},V_{i,j-1/2},V_{i,j+1/2})|},\right.\\ 
\left.\frac{|V_{i,j}-\min(V_{l,m})|}{|V_{i,j}-\min(V_{i-1/2,j},V_{i+1/2,j},V_{i,j-1/2},V_{i,j+1/2})|}\right), 
\end{multline} 
where $l=i-2,i-1,i,i+1,i+2$, $m=j-2,j-1,j,j+1,j+2$ except $(l,m)=(i,j)$. 
In a 3D case, the limiting function (\ref{es56}) must include all the neighbors 
of the cell $(i,j)$. 
 
\section{A FORTRAN implementation} 
 
The algorithm for computing the left and the right boundary values ${\bf V}^{L}$ and 
${\bf V}^{R}$ (\ref{eqxx21b})-(\ref{eqxx22b}) could be implemented in FORTRAN this way: 
 
\vspace{0.2cm} 
 
\noindent\verb"integer n, n2" 
 
\noindent\verb"real dt, dx" 
 
\noindent\verb"real VL(8), VC(8), VR(8), Vm(8), V(8), D(8), dVy(8)" 
 
\noindent\verb"real Lambda(8), sumL(8), VLnew(8), VLinterface(8)" 
 
\noindent\verb"real B(8,8), L(8,8), R(8,8)" 
 
\vspace{0.2cm} 
 
\noindent\verb"call Eigenvalues(VC,Lambda)" ! compute the eigenvalues at the center of the  
 
\noindent\hspace{5cm} ! cell $i$ (\verb"VC" - a state in the center) 
 
\noindent\verb"if (Lambda(1).gt.0.) then"\hspace{0.37cm} ! if $\lambda^1>0$ then 
compute a new state \verb"VLnew" 
 
\noindent\hspace{5cm} ! on the left side of the interface between the 
 
\noindent\hspace{5cm} ! cells $i$ and $i+1$ 
   
\noindent\verb"  VLnew=0."                          
 
\noindent\verb"  xi=1.- Lambda(1)*dt/dx"\hspace{0.56cm} ! $\xi$ in (\ref{eq1}) for $\lambda^1$ (see~(\ref{xieq})) 
 
\noindent\verb"  call Vxi(xi,VL,VC,VR,Vm)"\hspace{0.21cm} ! formula (\ref{eq1}); \verb"VL", \verb"VR" - 
the left and the right   
 
\noindent\hspace{5cm} ! boundary values of a parabola, \verb"Vm" - the result  
 
\noindent\verb"  call MatrixB(VC,B)"\hspace{1.3cm} ! compute the matrix $B$ 
 
\noindent\verb"  call Vectors(VC,L,R)"\hspace{0.96cm} ! compute the left (\verb"L") and right (\verb"R") eigenvectors 
 
\noindent\verb"  D=0."				 
 
\noindent\verb"  do n=1,8" 
 
\noindent\verb"    do n2=1,8" 
 
\noindent\verb"      D(n)=D(n)+B(n,n2)*dVy(n2)/dy"\hspace{0.2cm} ! $\left(B^{\,p}\:\partial_{\,y}V^p\right)$ (see (\ref{eqxx27c})-(\ref{Deq})) 
 
\noindent\verb"    enddo" 
 
\noindent\verb"  enddo"	 
 
\noindent\verb"  D=D*dt/2.   " 
 
\noindent\verb"  sumL=0." 
 
\noindent\verb"  do n=1,8" 
 
\noindent\verb"    if (Lambda(n).gt.0.) then"\hspace{0.2cm} ! only these waves affect the interface 
 
\noindent\verb"      xi=1.- Lambda(n)*dt/dx"\hspace{0.39cm}  ! $\xi$ in (\ref{eq1}) for $\lambda^p$ 
 
\noindent\verb"      call Vxi(xi,VL,VC,VR,V)"\hspace{0.2cm} ! \verb"V" - the result (a state at the point $\xi$) 
 
\noindent\verb"      do n2=1,8" 
 
\noindent\verb"        sumL(n)=sumL(n)+L(n,n2)*(V(n2)-Vm(n2)-D(n2))" ! a part of (\ref{eqxx21b}) 
 
\noindent\verb"      enddo" 
 
\noindent\verb"      do n2=1,8" 
 
\noindent\verb"        VLnew(n2)=VLnew(n2)+R(n2,n)*sumL(n)"  ! a part of (\ref{eqxx21b}) 
 
\noindent\verb"      enddo" 
 
\noindent\verb"    end if" 
 
\noindent\verb"  enddo" 
    
\noindent\verb"  VLnew=Vm+VLnew"\hspace{1.1cm} ! the result for (\ref{eqxx21b}) 
 
\noindent\verb"else"\hspace{3.31cm} ! if the maximum eigenvalue $\lambda^1<0$ 
 
\noindent\verb"  VLnew= VLinterface"\hspace{0.38cm} ! we keep the old value on the left side of the interface 
 
\noindent\verb"end if" 
\vspace{0.2cm} 
 
A FORTRAN code for computing $\overline{\vphantom{V^L}\,{\bf V}}^{\;L}$ and 
$\overline{\vphantom{V^R}\,{\bf V}}^{\,R}$ is similar. We only need to replace 
the function call \verb"call Vxi(xi,...)" with one that computes the integral (\ref{eq6}) 
or (\ref{eq7}), using  
 
\noindent\verb"xi=Lambda(1)*dt/dx" 
 
\noindent or 
 
\noindent\verb"xi=-Lambda(8)*dt/dx" ,
 
\noindent respectively. \verb"Lambda(8)" is the maximum absolute value of the negative eigenvalue. 
 
To obtain a monotonic solution for a 3D MHD problem, we suggest using the following algorithm: 
\begin{enumerate} 
\item Compute the average interface states $\overline{\vphantom{V^L}\,{\bf V}}^{\;L}$ 
and $\overline{\vphantom{V^R}\,{\bf V}}^{\,R}$ from (\ref{eqxx21z})-(\ref{eqxx22z}). 
\item Solve Riemann problem between $\overline{\vphantom{V^L}\,{\bf V}}^{\;L}$ 
and $\overline{\vphantom{V^R}\,{\bf V}}^{\,R}$ to determine the fluxes. 
\item Use a conservative difference scheme (\ref{eqxx2}) to compute new central states. 
\item Modify the magnetic field at the centers accordingly, see section~\ref{field}. 
\item Compute the interface states ${\bf V}^{L}$ and ${\bf V}^{R}$ from (\ref{eqxx21b})-(\ref{eqxx22b}). 
\item Solve Riemann problem between ${\bf V}^{L}$ and ${\bf V}^{R}$ to determine the new interface states. 
\item Apply procedure~1 in $x$-direction (section~\ref{mono}). 
\item Apply procedure~1 in $y$-direction (section~\ref{mono}). 
\item Apply procedure~1 in $z$-direction (section~\ref{mono}). 
\item Apply procedure~2 (section~\ref{mono}). 
\item Apply PPM procedure (\ref{eq4})-(\ref{eq5}). 
\end{enumerate} 
 
Note that only the interface states are modified by the monotonicity preserving procedures. 
The solution of the Riemann problem joins the interface values, but the monotonicity procedures 
split them again into the left and right values. 
 
\section{Numerical tests} 
 
\subsection{Riemann problem with multiple weak discontinuities} 

\begin{figure}[!h] 
\centering 
\includegraphics[width=120mm,height=50mm]{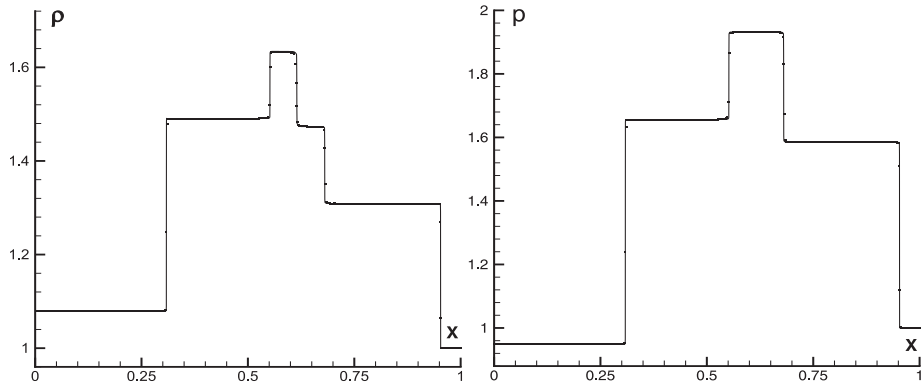} 
\caption{Riemann problem with multiple weak discontinuities. Density and pressure distributions.} 
\label{1D_1} 
\end{figure}

\begin{figure}[!h] 
\centering 
\includegraphics[width=120mm,height=50mm]{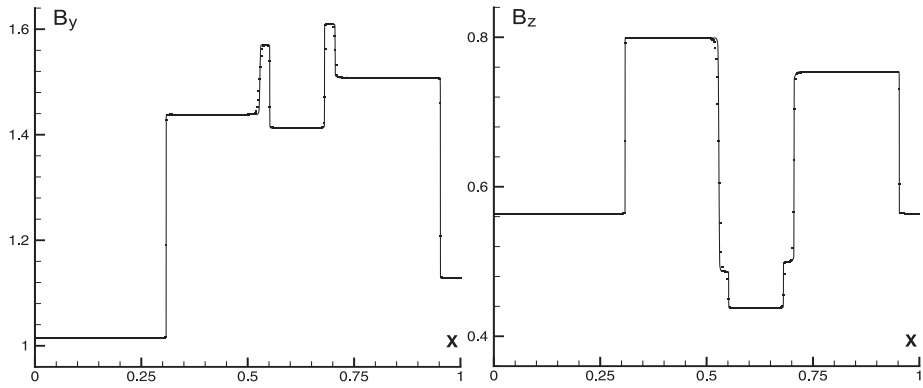} 
\caption{Same as in Fig.~\ref{1D_1} but for $y$- and $z$-components of the magnetic field.} 
\label{1D_2} 
\end{figure}

\begin{figure}[!h] 
\centering 
\includegraphics[width=120mm,height=50mm]{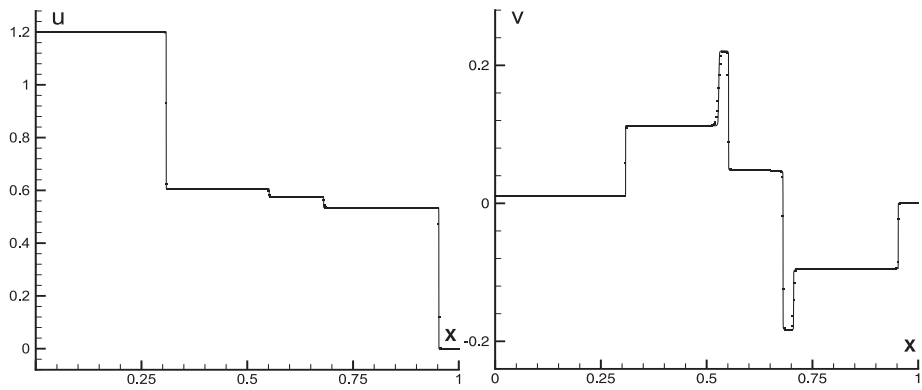} 
\caption{Same as in Fig.~\ref{1D_1} but for $x$- and $y$-components of the velocity.} 
\label{1D_3} 
\end{figure}

This is a 1D problem from~\cite{Dai_111}. An interval $x\in[0\ldots1]$ is divided 
in two by $x=0.5$. The left and the right states at the initial moment are defined as 
\begin{equation*} 
%\label{ex74} 
\left(\rho^{\,L},u^{\,L},v^{\,L},w^{\,L},B_y^{\,L},B_z^{\,L},p^{\,L}\right)=(1.08,1.2,0.01,0.5,3.6,2,0.95), 
\end{equation*} 
\begin{equation*} 
%\label{ex75} 
\left(\rho^{\,R},u^{\,R},v^{\,R},w^{\,R},B_y^{\,R},B_z^{\,R},p^{\,R}\right)=(1,0,0,0,4,2,1), 
\end{equation*} 
$B_x=2$, $\gamma=5/3$, $N=512$. The solution involves two fast shocks with Mach 
numbers 1.22 and 1.28, two slow shocks with Mach numbers 1.09 and 1.07, two rotational 
and one contact discontinuities. The solution for the moment $t=0.2$ is presented in 
Figs.~\ref{1D_1}-\ref{1D_3}. The solid line represents the exact solution and points
represent the numerical one. PPML produces very sharp fronts resolved with only a 
few grid points. 

 
\subsection{Numerical dissipation and decay of Alfv\'en waves} 
 
Numerical calculations on a discrete grid always lead to energy loss due to 
numerical dissipation. In order to estimate the properties of numerical 
dissipation of the PPML ideal MHD scheme, we used the test problem from~\cite{Ryu} 
and followed a decay of two-dimensional Alfv\'en wave. We used a standing wave 
propagating along the grid diagonal with initial conditions 
\begin{equation*} 
%\label{Rye1} 
\delta v_x=v_{amp}\:c_a\:\sin(k_xx+k_yy), 
\end{equation*} 
\begin{equation*} 
%\label{Rye2} 
\delta\rho=\delta p=\delta v_x=\delta v_y=\delta B_x=\delta B_y=\delta B_z = 0 
\end{equation*} 
in a stationary background flow with $\rho_0=1$, $p_0=1$, $B_x=1$, $B_y=B_z=0$. 
This gives the sound speed $c=1.291$ and the Alfv\'en velocity $c_a=0.7071$. 
The computational domain is a square box with size $L=1$ divided into $64\times64$ 
grid cells. The wavenumbers $k_x=k_y=2\pi/L$, the total wave number 
$k=\sqrt{k_x^2+k_y^2}=\sqrt 2(2\pi/L)$, the initial peak amplitude $v_{amp}=0.1$, 
adiabatic exponent $\gamma=5/3$. Computations were carried out with a Courant number 
$\sigma=0.4$. We used the periodic boundary conditions.
  
\begin{figure}[t] 
\centering 
\includegraphics[width=120mm,height=54.533mm]{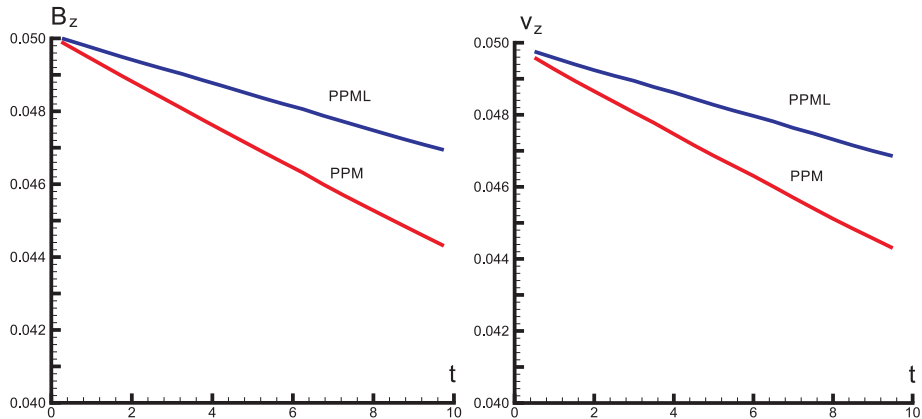} 
\caption{Decay of Alfv\'en waves. The maximum values of $B_z$ and $v_z$ as a functions of time.} 
\label{Alfven} 
\end{figure}

Figure~\ref{Alfven} shows the envelope for the maxima of $z$-component of the 
magnetic field and velocity obtained with PPML and PPM reconstruction procedures 
as a function of time. While both schemes show very low dissipation, PPML 
dissipation is even smaller than that of PPM. 
 
\subsection{Travelling circularly polarized Alfv\'en wave problem} 
 
This problem was suggested in~\cite{Toth} as a test for numerical accuracy of
smooth flow solutions. The circularly polarized Alfv\'en wave propagates at an 
angle of $\alpha=30^{{\rm o}}$ with respect to an axis $x$ in the domain 
$[0,1/\cos\alpha]\times[0,1/\sin\alpha]$. The initial conditions are 
\begin{equation*} 
\label{es61} 
\rho=1,\quad v_{\parallel}=0,\quad v_{\perp}=0.1\sin(2\pi\xi),\quad w=0.1\cos(2\pi\xi) 
\end{equation*} 
\begin{equation*} 
\label{es62} 
B_{\parallel}=1,\quad B_{\perp}=0.1\sin(2\pi\xi),\quad B_z=0.1\cos(2\pi\xi),\quad p=0.1, где 
\end{equation*} 
where $\xi=x\cos\alpha+y\sin\alpha$. For convenience the parallel and the orthogonal 
to the direction of Alfv\'en wave propagation components of the velocity and the 
magnetic field are used instead the of the components $u$, $v$, $B_x$ и $B_y$. 
For example $B_{\parallel}=B_x\cos\alpha+B_y\sin\alpha$, $B_{\perp}=B_y\cos\alpha-B_x\sin\alpha$. 
Alfv\'en wave travels to the point $(x,y)=(0,0)$ with the velocity $B_{\parallel}/\sqrt{\rho}=1$. 
Note that the wave becomes standing if $v_{\parallel}=1$. 
 
\begin{figure}[t] 
\centering 
\includegraphics[width=80mm,height=68.894mm]{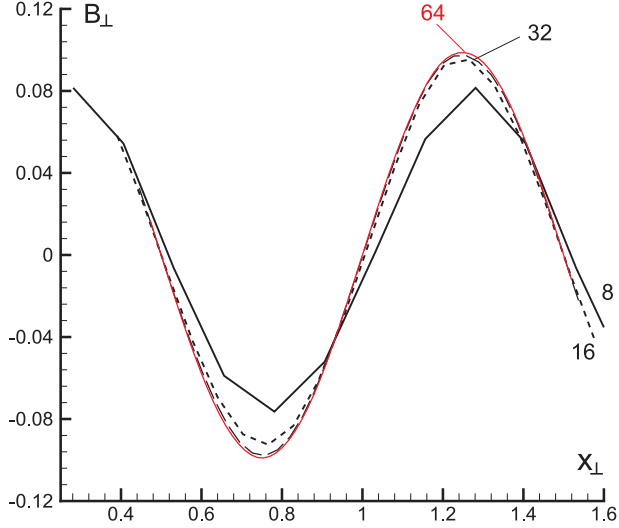} 
\caption{The orthogonal component $B_{\perp}$ in travelling Alfv\'en wave in computations on 
meshes with $N=8$, $16$, $32$ and $64$.} 
\label{Alfven2} 
\end{figure} 

\begin{table}[b] 
\caption{Travelling Alfv\'en wave. The average relative errors and the rates of convergence at $t=0.5$.} 
\begin{center} \footnotesize 
\begin{tabular}{|c|c|c|} 
\hline 
\rule{0pt}{12pt}\rule{6pt}{0pt}$N$\rule{6pt}{0pt} &\rule{25pt}{0pt} $\bar\delta_N$\rule{25pt}{0pt} &\rule{8pt}{0pt} $R_N$\rule{8pt}{0pt}\\ 
\hline 
8  & $2.2384\times10^{-1}$ & -    \\ 
16 & $5.7258\times10^{-2}$ & 1.967\\ 
32 & $1.7031\times10^{-2}$ & 1.755\\ 
64 & $5.0365\times10^{-3}$ & 1.771\\ 
\hline 
\end{tabular} 
\end{center}  
\label{table1}  
\end{table} 

The problem was solved on a set of rectangular $N\times2N$ meshes with $N=8$, $16$, 
$32$ and $64$. The averaged relative numerical errors were estimated as 
\begin{equation} 
\label{es63} 
\delta_N(U)=\frac{\sum\limits_{i=1}^N\sum\limits_{j=1}^{2N}|U_{i,j}^N-U_{i,j}^{E}|}{\sum\limits_{i=1}^N\sum\limits_{j=1}^{2N}|U_{i,j}^{E}|},\quad\mbox{for}\;U=v_{\perp}, w, B_{\perp}, B_z, 
\end{equation} 
where the solution on the mesh $N=128$ regarded as the exact one $U_{i,j}^{E}$. 
The rate of convergence was calculated as follows 
\begin{equation} 
\label{es64} 
R_N={\rm log}_2\left(\delta_{N/2}/\delta_N\right), 
\end{equation} 
where $\delta_N$ is an averaged value: 
\begin{equation} 
\label{es65} 
\delta_N=\frac14\left(\vphantom{r^A}\delta_N(v_{\perp})+\delta_N(w)+\delta_N(B_{\perp})+\delta_N(B_z)\right). 
\end{equation} 
 
\begin{figure}[t] 
\centering 
\includegraphics[width=120mm,height=58.51mm]{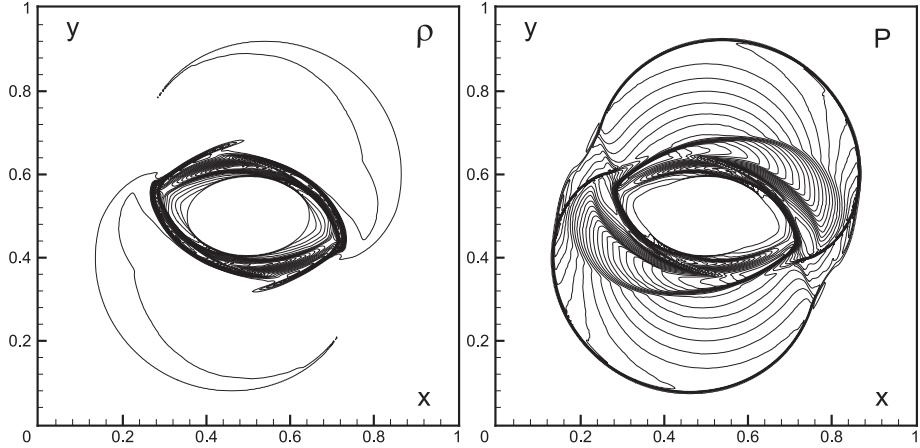} 
\caption{Rotor problem. The contours represent thirty levels of the density in 
the range from 1.3 to 13.5 and the pressure in the range from 0.12 to 2.1 at 
$t=0.15$.} 
\label{rotor1} 
\end{figure} 
\begin{figure}[!t] 
\centering 
\includegraphics[width=120mm,height=58.51mm]{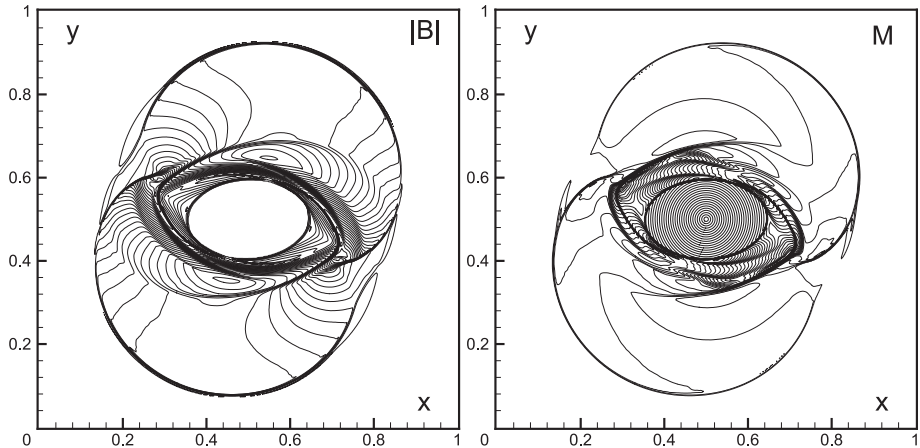} 
\caption{Rotor problem. The lines represent thirty levels of absolute value 
of the magnetic field in the range from 0.32 to 2.288 and the Mach numbers 
in the range between 0.144 and 4.27 at $t=0.15$.} 
\label{rotor2} 
\end{figure} 
 
The calculations were carried out up to $t=5$ with a Courant number $\sigma=0.4$ 
and $\gamma=5/3$. We applied periodic boundary conditions. 
Figure~\ref{Alfven2} demonstrates the 
convergence of the numerical solution. It shows the orthogonal component $B_{\perp}$ in 
travelling Alfv\'en wave for computations on meshes of different size $N$. 
Table~\ref{table1} gives average relative numerical errors and rates of 
convergence obtained for the PPML scheme.  
 
\subsection{Rotor problem} 

The Rotor problem was suggested in~\cite{Balsara_S} and has been widely used to test 
numerical schemes~\cite[e.g.,][]{Toth,Han,Flash}. It turned out to be a hard nut 
to crack for many codes due to the appearance of negative pressure values~\cite{Toth}. 
 
The computational domain in this case is a square $[0,1]\times[0,1]$ with a uniform pressure $p=1$ 
and magnetic field components $B_x=5/\sqrt{4\pi}$, $B_y=0$. There is a rotating disk of dense 
fluid at the center with a radius $r_0=0.1$. For $r<r_0$ we specify $\rho=10$, $u=-v_0(y-0.5)/r_0$, 
$v=v_0(x-0.5)/r_0$, where $r=\sqrt{(x-0.5)^2+(y-0.5)^2}$, $v_0=2$. For $r>r_1=0.115$ the fluid is 
initially at rest ($u=v=0$) with density $\rho=1$. In the intermediate zone $r_0<r<r_1$ we use a 
linear interpolation of the variables: $\rho=1+9f$, $u=-fv_0(y-0.5)/r$, $v=fv_0(x-0.5)/r$, 
$f=(r_1-r)/(r_1-r_0)$. In this setup, the initial configuration is imbalanced due to centrifugal 
forces. The rotating fluid will tend to equilibrate, while the magnetic field holds the oblate 
shape of the rotor. 
 
The computations were carried out on a set of $N\times N$ meshes with $N=50$, 100, 200 and 400, 
with a Courant number $\sigma=0.4$ and $\gamma=1.4$ until $t=0.15$. The boundary conditions are 
obtained through zero-order interpolation. Figures~\ref{rotor1} and \ref{rotor2} show the flow fields
for $N=400$.  
 
\begin{table}[!h] 
\caption{The rotor problem. The average relative errors and the rates of convergence in the 
numerical codes at $t=0.15$.} 
\begin{center} \footnotesize 
\begin{tabular}{|c|c|c|c|c|} 
\hline 
\multicolumn{1}{|c|}{\rule{0pt}{10pt}N}&\multicolumn{2}{c|}{PPML}&\multicolumn{2}{c|}{Flash3 USM-MEC}\\ 
\cline{2-5} 
\rule{0pt}{10pt}&\multicolumn{1}{c|}{$\bar\delta_N$}& \multicolumn{1}{c|}{$R_N$}&\multicolumn{1}{c|}{$\bar\delta_N$}&\multicolumn{1}{c|}{$R_N$}\\ 
\hline 
50\rule{0pt}{12pt} & $9.4274\times10^{-2}$ & - & $1.1470\times10^{-1}$ & $ - $\\ 
\rule{6pt}{0pt}100\rule{6pt}{0pt} & \rule{6pt}{0pt}$4.5204\times10^{-2}$\rule{6pt}{0pt} &\rule{6pt}{0pt}1.06 \rule{4pt}{0pt}&\rule{6pt}{0pt} $5.9800\times10^{-2}$\rule{6pt}{0pt} &\rule{6pt}{0pt} 0.94\rule{6pt}{0pt}\\ 
200 & $1.9262\times10^{-2}$ & 1.24 & $2.5000\times10^{-2}$ & 1.26\\ 
\hline 
\end{tabular} 
\end{center}  
\label{table2}  
\end{table}

We compare numerical solutions obtained with PPML with those presented in~\cite{Toth,Han,Flash}. 
In Table~\ref{table2} the average relative numerical errors and the rates of convergence 
are given for PPML and for Flash3 USM-MEC (unsplit staggered mesh algorithm with modified electric 
field construction introduced in~\cite{Flash}). Both PPML and Flash3 USM-MEC codes use a Roe 
solver~\cite{Roe} for this test. The relative numerical errors were computed using 
eq.~(\ref{es63}), where for $U_{i,j}^E$ we used the highest resolution result ($N=400$). 
The average error $\bar\delta_N(U)$ is defined as the average $\delta_N(U)$ for all non-zero 
variables $U$. The rate of convergence is estimated as in~(\ref{es64}). PPML results are more 
accurate and have a comparable rate of (self-)convergence with those from the new Flash3 MHD 
solver. 
 
\begin{figure}[t] 
\centering 
\vspace{-1.2cm}
\includegraphics[scale=0.6]{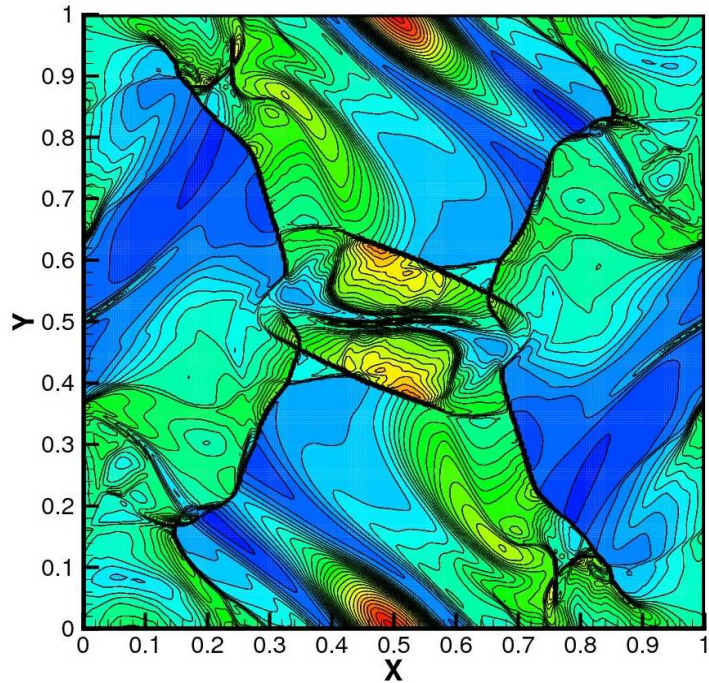} 
\vspace{-0.5cm}
\caption{Orszag-Tang vortex problem. The contours represent thirty levels of 
pressure equally spaced in the range from 0.02 to 0.5 at $t=0.5$.} 
\label{test4_2D} 
\end{figure} 
\begin{figure}[!t] 
\centering 
\vspace{-0.4cm}
\includegraphics[scale=0.8]{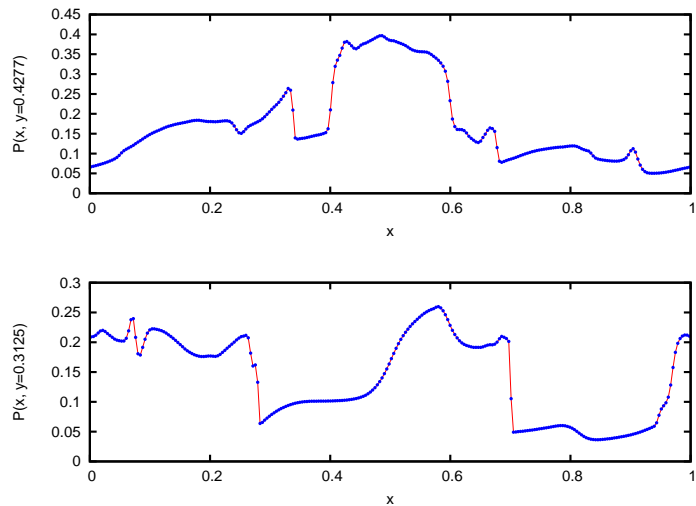}
\caption{Orszag-Tang vortex problem. The pressure along the lines $y=0.3125$ and $y=0.4277$ at $t=0.5$.} 
\label{test4_1D} 
\end{figure} 
 
\subsection{Orszag-Tang vortex problem} 
 
This problem was suggested in~\cite{Orszag} and since then has been used in many 
papers as a standard test problem for numerical codes in 2D MHD. It involves formation 
and an interaction of multiple shocks and a transition to supersonic turbulence.  
 
In the computational domain $[0,1]\times[0,1]$, we set a uniform density $\rho=25/(36\pi)$ 
and pressure $p=5/(12\pi)$ with $\gamma=5/3$ (in this case the sound velocity 
$c=\sqrt{\gamma p/\rho}=1$). The initial velocities and components of the 
magnetic field are set using harmonic functions: $u=-\sin2\pi y$, $v=\sin2\pi x$, $w=0$, 
$B_x=-B_0\sin2\pi y$, $B_y=B_0\sin4\pi x$, $B_z=0$, where $B_0=1/\sqrt{4\pi}$. Despite such 
smooth initial conditions the fluid motion becomes very complex. 
 
We carried out computations on $N\times N$ meshes with several values of $N$ using periodic 
boundary conditions and a Courant number $\sigma=0.3$. Figure~\ref{test4_2D} demonstrates the 
pressure distribution at time $t=0.5$ for $N=256$. In Fig.~\ref{test4_1D} the pressure distributions 
along the lines $y=0.3125$ ($j=83$) and $y=0.4277$ ($j=112$) are shown to illustrate the 
accuracy and sharpness of the main flow features. 
 
Table~\ref{table3} contains the average relative numerical errors and the rates of 
convergence for PPML and Flash3 USM-MEC~\cite {Flash} solvers at $t=0.5$. As an ``exact'' 
solution $U_{i,j}^E$, we used a solution obtained on the grid with $N=400$. The corresponding 
PPML results are more accurate and demonstrate better convergence. 
\begin{table}[h] 
\caption{Orszag-Tang vortex problem. The averaged relative errors and the rates of convergence 
in the numerical codes at $t=0.5$.} 
\begin{center} \footnotesize 
\begin{tabular}{|c|c|c|c|c|} 
\hline 
\multicolumn{1}{|c|}{\rule{0pt}{10pt}N}&\multicolumn{2}{c|}{PPML}&\multicolumn{2}{c|}{Flash3 USM-MEC}\\ 
\cline{2-5} 
\rule{0pt}{10pt}&\multicolumn{1}{c|}{$\bar\delta_N$}& \multicolumn{1}{c|}{$R_N$}&\multicolumn{1}{c|}{$\bar\delta_N$}&\multicolumn{1}{c|}{$R_N$}\\ 
\hline 
50\rule{0pt}{12pt} & $8.9095\times10^{-2}$ & - & $1.0160\times10^{-1}$ & $ - $\\ 
\rule{6pt}{0pt}100\rule{6pt}{0pt} & \rule{6pt}{0pt}$4.4249\times10^{-2}$\rule{6pt}{0pt} &\rule{6pt}{0pt}1.013 \rule{6pt}{0pt}&\rule{6pt}{0pt} $5.2200\times10^{-2}$\rule{6pt}{0pt} &\rule{6pt}{0pt} 0.964\rule{6pt}{0pt}\\ 
200 & $1.8851\times10^{-2}$ & 1.235 & $1.9900\times10^{-2}$ & 1.390\\ 
\hline 
\end{tabular} 
\end{center}  
\label{table3}  
\end{table}

\section{A compressible turbulence simulation} 
Magnetized supersonic turbulence plays an important role in statistical star formation
theories~\cite{mckee.07}. This stimulated development of accurate numerical methods suitable for modeling
turbulent molecular clouds. One of the motivations behind the design of PPML has 
been a need for an MHD scheme with low numerical dissipation comparable or better than 
that of PPM. In this section we illustrate the performance of PPML on a 
challenging problem of forced super-Alfv\'enic turbulence. Some numerical methods 
that successfully pass the tests discussed above turn unstable on this 
application. Since adding more dissipation where needed
-- the usual way to cure for ``blow ups'' caused by numerical instabilities -- would 
ultimately damage the derived statistics of turbulence \cite{yee.07}, the
issue of inherent stability of numerical methods is crucial for both supersonic
turbulence and star formation simulations.

\begin{figure}[!t] 
\centering 
\includegraphics[scale=0.7]{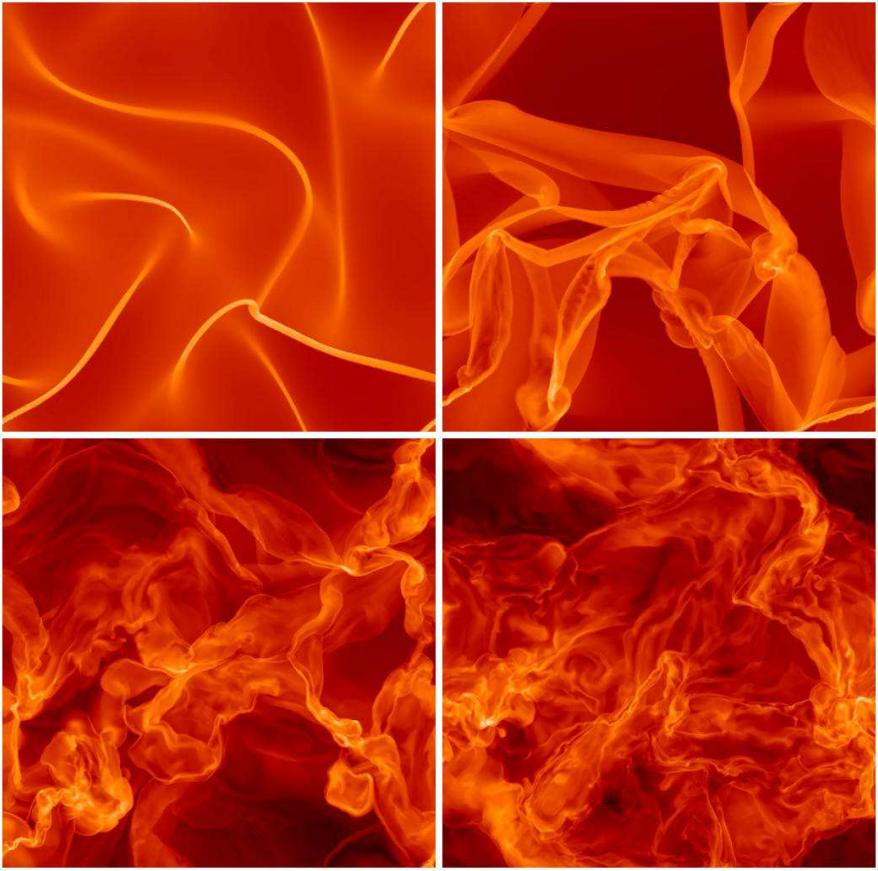} 
%\vspace{-0.1cm}
\caption{Supersonic turbulence simulation with PPML on a $512^3$ grid.
Four snapshots show the density field on a slice $x\equiv0$ illustrating 
a transition to fully developed turbulence with $M_s\approx M_A\approx10$. 
The transition includes formation of first strong shocks on the caustics 
of the initial solenoidal velocity field ($t=0.5t_d$, top-left), shock 
interactions and active development of first shear instabilities ($t=1t_d$, top-right),
and a gradual transition to a statistical steady state  ($t=2t_d$ and $4t_d$, bottom row).
The standard logarithmic black-red-white color ramp shows high-density regions in 
light-red and rarefactions in black.} 
\label{turbo} 
\end{figure} 

For illustrative purposes, we present here a simulation of weakly magnetized supersonic 
turbulence. In this experiment, turbulence in a periodic domain of linear dimension 
$L=1$ is driven by a large-scale solenoidal force for 8 flow-crossing times 
$t_d\equiv L/2M_s$. At time $t=0$, a uniform gas with density $\rho\equiv1$ is 
permeated by a weak uniform magnetic field ${\bf B}_0\parallel\pmb{x}$, such that 
$\beta_0\equiv 2p/B_0^2=20$. We apply an initial large-scale 
velocity field that corresponds to an rms sonic Mach number $M_s\sim10$ and assume 
an isothermal equation of state ($c\equiv1$) to mimic the average physical conditions 
in the dense parts of molecular clouds ($n=10^3$~cm$^{-3}$, $T=10$~K).

\begin{figure}[!t] 
\vspace{-0.1cm}
  \begin{center}
    \mbox{
      \subfigure{\epsfig{file=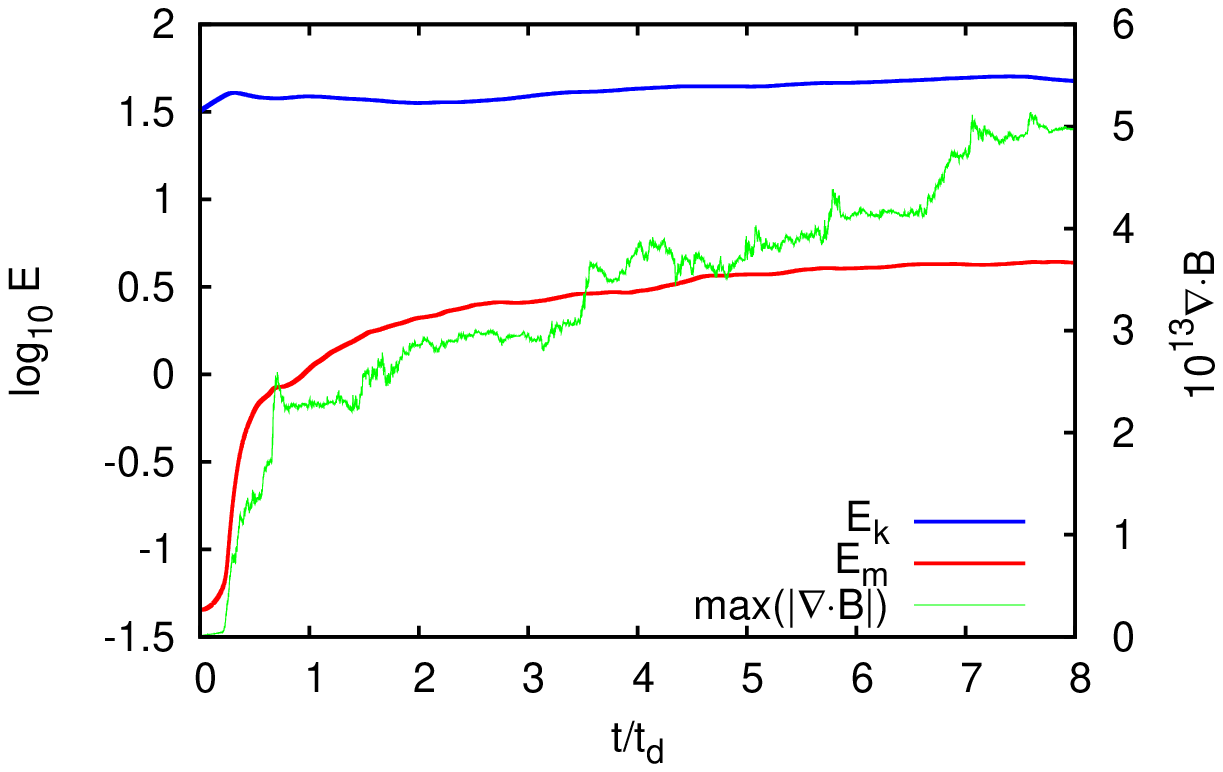,scale=0.53}} \quad
\vspace{-0.1cm}
      \subfigure{\epsfig{file=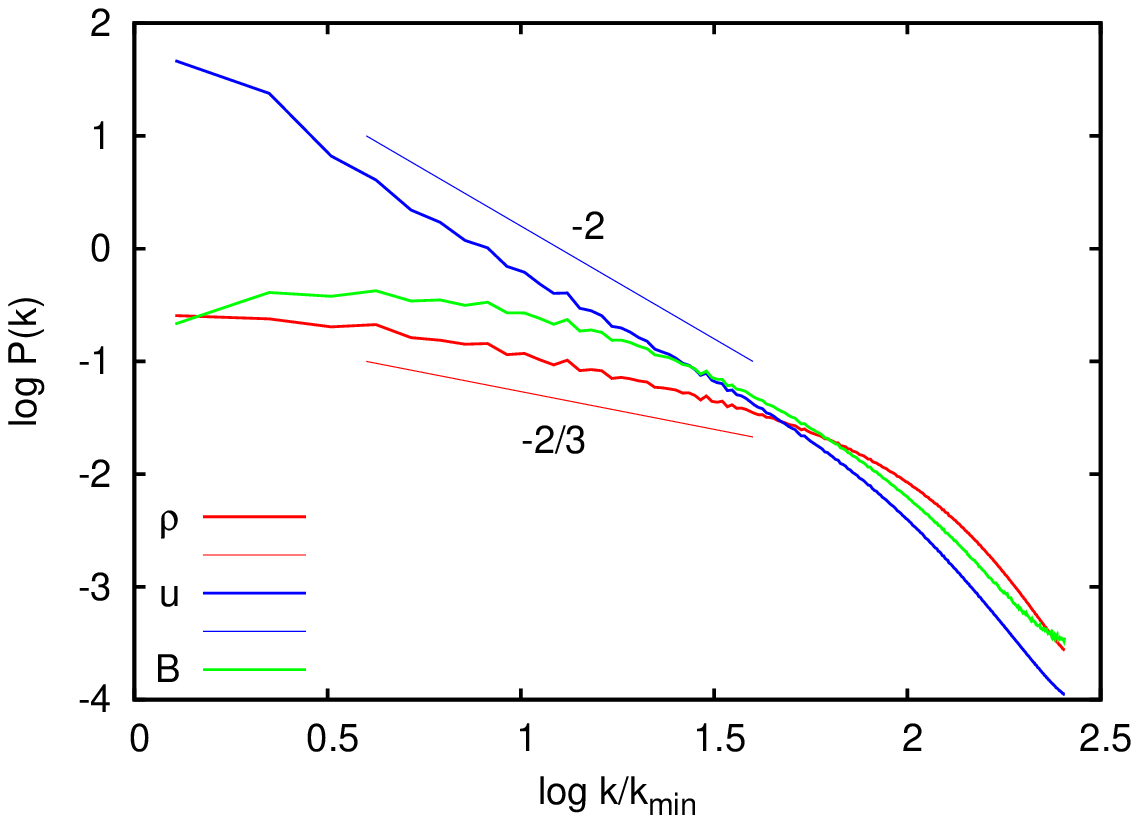, scale=0.50}}
      }
  \end{center}
\vspace{-1cm}
    \caption{Supersonic turbulence simulation with PPML on a $512^3$ grid. Time-evolution of
kinetic and magnetic energy and maximum absolute value of $\pmb{\nabla\cdot}{\bf B}$ (left panel)
and turbulent power spectra for the velocity, density and magnetic field (right panel).}
\label{turbo2} 
\end{figure}

The evolution begins with the formation of strong shocks on ``caustics'' of the initial velocity
field. Shock interactions then cascade the initial kinetic energy of large-scale motion of the 
gas down to smaller and smaller scales \'a la Kolmogorov-Richardson, see Fig.~\ref{turbo}. 
The magnetic field gets amplified by a factor of about 50 via the small-scale dynamo action
\cite{schekochihin.07}. 
The large-scale solenoidal force (acceleration) keeps the rms sonic Mach number roughly 
constant at $M_s\approx10$. 
The evolution of kinetic and magnetic energies is shown in Fig.~\ref{turbo2}, 
left panel. Also shown is $max(\left|\pmb{\nabla\cdot}{\bf B}\right|)$ as a function of time during 
this simulation. The method keeps the absolute value of the divergence of magnetic field 
below $10^{-12}$ at all times, even after 70,000 integration time steps (if double precision 
is used). After about 4 crossing times of evolution, the 
system completes a transition to a fully developed isotropic state with 
$M_s\approx10$ and $M_A\approx10$. The right panel of Fig.~\ref{turbo2} illustrates spectral 
characteristics of this saturated state by showing the time-average (over 25 flow snapshots 
taken between $t=4t_d$ and $t=8t_d$) power spectra for the density, velocity, and 
magnetic field strength.

The velocity spectrum has an extended scaling range with a slope of about $-2$, as
in the Burgers turbulence, similar to the corresponding scaling in non-magnetized flows at
high Mach numbers \cite{kritsuk...07}. This is expected, as turbulence here is only
weakly magnetized. The density spectrum slope of about $-0.7$ is
again consistent with our previous results for non-magnetized flows obtained with PPM
($-1.0$ at $M_s=6$) and with an anticipated trend towards a flat ``white noise'' 
spectrum at $M_s=\infty$. 
The magnetic energy spectrum does not show a clear scaling range, as expected at 
this modest resolution, assuming the effective magnetic Prandtl number of PPML is 
of order unity. We also looked at more advanced spectral characteristics for 
compressible flows, such as the power spectrum of $\rho^{1/3}{\bf v}$, and found a slope 
of $-1.7$. This power spectrum related to the energy transfer rate
in wavenumber space is insensitive to the turbulent Mach number and
should have a Kolmogorov $-5/3$ slope \cite{kritsuk...07} in both incompressible
and highly compressible regimes, although a steeper scaling does occur due to 
intermittency \cite{kritsuk...07b}. 
This is true for both non-magnetized and weakly magnetized flows.

We have also carried out two additional simulations of the same kind but with higher
degrees of magnetization, $\beta_0=2$ and 0.2 \cite{kritsuk...08}. While the saturated
turbulent state in the $\beta_0=2$ variant is still super-Alfv\'enic with $M_A\approx3$ 
and the magnetic energy is about 3 times smaller than the kinetic energy, 
the trans-Alfv\'enic case, $\beta_0=0.2$, reaches an equipartition of kinetic and magnetic 
energies. In both cases, PPML proved to be perfectly stable at a Courant number $\sigma=0.2$, 
as in the super-Alfv\'enic case $\beta_0=20$ discussed above. 

Our approach to handle the stability issues in MHD turbulence simulations 
with PPML is as follows: (i) we use locally multidimensional reconstruction 
that improves the quality of Right and Left interface states and helps to 
avoid numerous well-known pathologies, such as ``carbuncles'', etc.~\cite{quirk94}; (ii) we control 
the quality of these states {\em before} moving forward with the flux calculation; 
(iii) if we find that the states are not satisfactory, we reduce the order of
reconstruction to linear or further skip the whole reconstruction step;
(iv) we use only nonlinear Riemann solver in all cases lacking the
reconstruction step. 

Overall, the derived spectral properties of weakly magnetized highly compressible 
turbulence demonstrate that low dissipation and wide spectral bandwidth of PPML 
make it an ideal numerical scheme for large-scale simulations of magnetized supersonic 
turbulence.

\section{Conclusions} 

In this paper we presented PPML, a new numerical method for compressible ideal MHD that 
is based on the piecewise parabolic approximation. Interface values for the interpolation 
parabolae in every grid cell are defined with the help of the Riemann invariants which 
remain constant along the characteristics. The monotonicity of the states on the interfaces 
between adjacent cells is provided by the monotonicity- and extrema-preserving procedure 
from~\cite{Rider}. The scheme is fully multidimensional as it includes the terms corresponding 
to the tangential directions in the amplitude equations. This helps to avoid numerous
well-known pathologies, such as ``carbuncles'', etc.

The states in the cell centers are defined by the conservative difference 
scheme (\ref{eqxx2}). To obtain the second-order temporal accuracy we must average 
the wave amplitudes over the corresponding domains of influence. To define the 
fluxes we need to solve the Riemann problem between the states at the cell interfaces
computed with the averaged amplitudes. 
 
To preserve zero divergence of the magnetic field in three dimensions, we 
use an unsplit Godunov method based on the constrained transport approach~\cite{Gardiner}. 
We use the information about the magnetic field gradients to fulfill the constraint 
on the magnetic field more accurately.

We tested the performance of PPML on several numerical problems which demonstrated
its high accuracy on both smooth and discontinuous solutions. Two-dimensional 
flow fields generated by PPML are highly resolved without any wiggled contour lines. 
Our pilot simulations of supersonic magnetized turbulence in three dimensions with PPML 
show that low dissipation and wide spectral bandwidth of this method make it an ideal 
candidate for direct turbulence simulations.

\section*{Acknowledgement}

This research was supported in part by the National Science Foundation through 
grants AST-0607675 and AST-0808184, as well as through TeraGrid resources 
provided by NICS, ORNL, SDSC, and TACC (allocations MCA-07S014 and MCA-98N020).

\end{document}